\documentclass[lettersize,journal]{IEEEtran}
\usepackage[misc,geometry]{ifsym}
\usepackage[dvipsnames]{xcolor}
\usepackage{graphicx}
\usepackage{multirow}
\usepackage{booktabs}
\usepackage{comment}
\usepackage{url}
\usepackage{balance}
\usepackage{enumitem}
\usepackage{hyperref}
\usepackage{cite}
\usepackage{alltt}
\usepackage{caption}
\usepackage{subcaption}
\usepackage{hyperref}

\begin{document}

\newcommand\mypound{\scalebox{0.8}{\raisebox{0.4ex}{\#}}}
\newcommand{\gray}[1]{\textcolor{Gray}{#1}}
\newcommand{\blue}[1]{\textcolor{NavyBlue}{#1}}
\newcommand{\red}[1]{\textcolor{Red}{#1}}
\newcommand{\bb}[1]{\mathbb{#1}}
\newcommand{\ttt}[1]{\textbf{\texttt{\small{#1}}}}

\title{The Closed Resolver Project: \\ Measuring the Deployment of Inbound Source Address Validation}

\author{Yevheniya Nosyk, Maciej Korczy\'nski, Qasim Lone, Marcin Skwarek, Baptiste Jonglez, Andrzej Duda
\thanks{Y. Nosyk, M. Korczy\'nski, A. Duda are with Univ. Grenoble Alpes, CNRS, Grenoble INP, LIG, F-38000 Grenoble, France.}
\thanks{Q. Lone is with RIPE NCC, the Netherlands.}
\thanks{M. Skwarek was with Univ. Grenoble Alpes, CNRS, Grenoble INP, LIG, F-38000 Grenoble, France. He is now with Exatel, Poland.}
\thanks{B. Jonglez was with Univ. Grenoble Alpes, CNRS, Grenoble INP, LIG, F-38000 Grenoble, France. He is now with Inria, France.}}
\markboth{IEEE/ACM TRANSACTIONS ON NETWORKING}
{Nosyk \MakeLowercase{et al.}: The Closed Resolver Project: Measuring the Deployment of Source Address Validation of Inbound Traffic}

\maketitle

\begin{abstract}
Ingress filtering, commonly referred to as Source Address Validation (SAV), is a practice aimed at discarding packets with spoofed source IP addresses at the network periphery. \textit{Outbound} SAV, i.e., dropping traffic with spoofed source IP addresses as it leaves its source network, has received widespread attention in operational and research communities. It is one of the most effective ways to prevent Reflection-based Distributed Denial-of-Service (DDoS) attacks. Contrariwise, \textit{inbound} SAV, i.e., dropping incoming spoofed traffic at the destination network edge, has received less attention, even though it provides protection for the deploying network. In this paper, we present the results of the Closed Resolver Project, our initiative aimed at finding networks without inbound SAV and raising awareness of the issue. We perform the first Internet-wide active measurement study to enumerate networks that enforce (or not) inbound SAV. We reach open and closed Domain Name System (DNS) resolvers in tested networks and determine whether they resolve requests with spoofed source IP addresses. Our method provides unprecedented insight into \textit{inbound} SAV deployment by network operators, revealing 49\% IPv4 and 26\% IPv6 Autonomous Systems (AS) that suffer from a consistent or partial absence of inbound filtering. By identifying dual-stack DNS resolvers and ASes, we further show that inbound filtering is generally deployed consistently across IPv4 and IPv6. Finally, the lack of inbound SAV exposes 2.5M IPv4 and 100K IPv6 purportedly closed DNS resolvers to many types of external attacks, including NXNSAttack, zone poisoning, or zero-day vulnerabilities in DNS software. 
\end{abstract}

\begin{IEEEkeywords}
IP spoofing, ingress filtering, Source Address Validation, DNS resolvers, IPv6, dual-stack.
\end{IEEEkeywords}

\IEEEpeerreviewmaketitle

\section{Introduction}

\IEEEPARstart{T}{he} Internet relies on IP packets to enable communication between hosts with the destination and source addresses specified in packet headers. However, there is no packet-level authentication mechanism to ensure that the source address has not been altered~\cite{Beverly:2009:UED:1644893.1644936}. The modification of a source IP address is referred to as ``IP spoofing''. It results in the anonymity of the sender and prevents a packet from being traced to its origin. Reflection-based Distributed Denial-of-Service (DDoS) attacks leverage this mechanism and become even more destructive using an amplification vector~\cite{hell, Kuhrer:2014:EHR:2671225.2671233, unchained}. As it is not possible in general to prevent packet header modification, concerted efforts have been undertaken to ``prohibit attackers from  using forged source addresses which do not reside within a range of legitimately advertised prefixes"~\cite{Ferguson:2000:NIF:RFC2827}. Ingress filtering, implemented at the periphery of Internet-connected networks, can achieve this goal. It was first formalized in RFC~2827, also known as Best Current Practice 38~\cite{Ferguson:2000:NIF:RFC2827}. In the network operator and research communities, ``ingress filtering" is commonly referred to as ``Source Address Validation (SAV)''~\cite{tracefilter} and the network ``periphery'' as the network ``edge". We adopt this widely-accepted terminology throughout this paper.

One reason why the volume of DDoS attacks is growing exponentially~\cite{googleddos} is the lack of \textit{outbound} SAV: it enables attackers to launch destructive attacks while staying untraceable. Some of the largest DDoS attacks known to date relied on sending spoofed requests to open Connection-less Lightweight Directory Access Protocol (CLDAP)~\cite{googleddos,aws}, DNS~\cite{googleddos,spamhaus}, Memcached~\cite{github}, and Simple Mail Transfer Protocol (SMTP)~\cite{googleddos} services. Preventing this type of attacks requires two actions. The first is to decrease the amplifier landscape by ensuring that, wherever possible, services do not accept requests from any host on the Internet. The second, and the most important one, is to deploy \textit{outbound} SAV so that all traffic leaving the attacker network can only have a valid source IP addresses. Given the prevalent role of IP spoofing in cyberattacks, the deployment of \textit{outbound} SAV was extensively studied in literature \cite{Kuhrer:2014:EHR:2671225.2671233,mauch,loops,Lichtblau,Lucas,qasimSP}.

However, little research investigated \textit{inbound} SAV, i.e., filtering out spoofed traffic at the edge of the destination network. The lack of \textit{inbound} SAV enables an external intruder to masquerade as an internal host of a network. In this case, it is enough for an attacker to spoof its source IP address so that it matches the destination network prefix. Such a spoofed request will reach the intended target (e.g., a closed DNS resolver only processing requests originating from its local network), even though the packet itself is coming from \textit{outside} the destination network. It may raise serious problems for the victim network: all the services that were believed to be only accessible to local clients are now accepting requests from attackers.

For example, network administrators may wish to secure DNS resolvers by making them closed. A closed resolver only accepts DNS queries from known clients and does so by matching the source IP address of a query with a list of allowed addresses. If the attacker spoofs any address from the list of allowed clients, she can launch any attack that commonly targets open resolvers and nameservers, such as zone poisoning~\cite{Korczynski:2016:ZPN:2987443.2987477}, cache poisoning~\cite{kaminsky}, DNS Unchained~\cite{unchained}, NXDOMAIN~\cite{LuoWXCYT18}, or NXNSAttack~\cite{NXNSAttack}. These attacks can result in Denial-of-Service for both recursive resolvers and authoritative nameservers, with a maximum packet amplification factor of 1,620 for NXNSAttack~\cite{NXNSAttack}. Although IP spoofing is not strictly required for these attacks to succeed, it can greatly increase the number of affected DNS resolvers and authoritative nameservers. Furthermore, IP spoofing allows targeting closed servers for which administrators do not expect such attacks, so, they are less likely to deploy preventive counter-measures or monitor these servers for attacks. Deploying inbound SAV at the network edge is an effective way of protecting closed DNS resolvers (and other services) from this type of external attacks. Therefore, it directly protects hosts inside the network, providing direct and immediate security benefits to the organization deploying it.

Unlike \textit{outbound} SAV, only a few initiatives (the Spoofer project~\cite{Spoofer} and a concurrent work by Deccio~et al.~\cite{behind}) measured the \textit{inbound} SAV deployment. In this article, we present results of the Closed Resolver Project~\cite{closed}---the first Internet-wide scanning campaign that enumerates networks not deploying inbound SAV. We extend our previous work~\cite{korczyski2020dont} and make the following main contributions:

\textit{1) We exhaustively enumerate networks that do not deploy inbound SAV for IPv4:} We propose a new measurement technique to identify networks that do not filter inbound traffic based on source IP addresses. We perform Internet-wide scans of all the routable IPv4 BGP prefixes collected by RouteViews~\cite{routeviews}. We send a DNS request of type \texttt{A} to each routable IP address (target address) in a packet with a spoofed source IP address. When sending the request to $X$, we choose $X+1$ as the source IP address belonging to the same prefix. If there is no filtering both in transit networks and at the destination network edge, the target will receive our request. If it is a DNS resolver and our spoofed address matches the list of allowed clients, the resolver will resolve our request. As we spoof the source IP address, the response from the resolver is not routed back to our scanner, preventing us from analyzing it. However, we control the authoritative nameserver for queried domains and we observe queries sent either directly by the resolver under test or through a chain of forwarding resolvers. Overall, this method identifies networks that do not correctly filter \textit{incoming} packets without the need for a vantage point inside the network itself. The only requirement is that the network contains a DNS resolver (possibly closed).

\textit{2) We enumerate networks that do not deploy inbound SAV in IPv6}: The adoption of IPv6 is gradually increasing~\cite{google6}, so the IPv6 Internet is becoming an attractive attack vector, especially considering that network operators do not protect the IPv6 portion of their networks as well as they do for IPv4~\cite{back_door}. Given the number of available addresses, a complete scan of the IPv6 address space (as explained previously for IPv4) is not computationally feasible. Gasser et al.~developed the IPv6 Hitlist Service~\cite{hitlist} that collects responsive IPv6 addresses from different sources, such as domain lists, DNS \texttt{ANY} lookups, Certificate Transparency (CT) logs, zone transfers~\cite{rfc5157,AXFR}, etc. To enrich this list, we also deploy a two-level DNS zone setup that forces resolvers to switch from IPv4 to IPv6 to resolve our domain names, thus discovering IPv6 resolvers as a by-product of the IPv4 scan. Then, we perform a scan of the enumerated IPv6 addresses using the same method as for IPv4.

\textit{3) We enumerate networks that deploy inbound SAV for IPv4 and IPv6}: The above technique, when applied alone, reveals the \emph{absence} of inbound SAV at the network edge. However, we can confirm the \emph{presence} of inbound SAV (possibly in transit) by following each spoofed query with a non-spoofed one to detect open resolvers. If servers reply to unspoofed requests but not to spoofed ones, we can infer the presence of SAV for incoming traffic either at the network edge or in transit networks. 
 
\textit{4) We combine different methods to check SAV compliance in both directions:} We collect the Spoofer data (over one month) and use a method proposed by Mauch~\cite{mauch} to infer the absence and the presence of \textit{outbound} SAV. In this way, we study the SAV deployment policies per provider in both directions. Previous work demonstrated the difficulty in incentivizing providers to adopt filtering for outbound traffic as it benefits other networks and not the network doing the deployment \cite{spoofer_new,qasimSP}. This work shows that even though SAV for inbound traffic directly benefits the networks implementing it, it is less widely deployed than outbound SAV.

\textit{5) We compare SAV deployment status for IPv4 and IPv6:} We first perform the analysis at the individual host level by identifying candidate dual-stack DNS resolvers. We configure our two-level DNS zones that require traversal from IPv4 to IPv6 (or the other way round). If a recursive DNS resolver has the connectivity over both versions of the IP protocol, we would see its IPv4 and IPv6 addresses on our authoritative nameserver to resolve a single domain name. For every such (IPv4, IPv6) address pair, we gather DNS-level information (\texttt{version.bind} and \texttt{PTR} records) and use other general-purpose fingerprinting tools to identify services running on ports 22, 80, 123, 443, and 587. Hardware and software information about each pair provides evidence on whether the two addresses belong to the same host or not. We then compare the filtering policies at the autonomous system level by analyzing ASes revealed during both IPv4 and IPv6 scans. As a result, we show that inbound SAV tends to be consistently deployed for IPv4 and IPv6.

\textit{6) We analyze the geographical distribution of networks vulnerable to spoofing of inbound traffic:} We show that the absence of inbound SAV is widespread and not limited to any particular geographic region. Furthermore, national Computer Security Incident Response Teams (CSIRTs) may find it valuable to know the extent to which networks under their governance are vulnerable to spoofing of inbound traffic and possibly cooperate in the notification campaign.

The rest of the paper is organized as follows. Section~\ref{sec:background} provides background on Source Address Validation and how it protects from spoofing attacks. Section~\ref{sec:related} discusses related work in the field. Section~\ref{sec:methodology} introduces our methodology, while Section~\ref{sec:results} provides scan results. Sections~\ref{sec:network}, \ref{sec:policies}, \ref{sec:deploymv4v6} discuss network characteristics, outbound vs. inbound SAV policies and IPv4 vs. IPv6 deployment, respectively. Section~\ref{sec:geo} discusses the geographic location of vulnerable networks. Section~\ref{limitations} details limitations and ethical considerations of the proposed method. Finally, Section~\ref{sec:conclusion} concludes the paper and gives some directions for future work.

\section{Background}\label{sec:background}

\begin{figure}[!t]
    \centering
    \begin{subfigure}[b]{0.49\textwidth}
        \includegraphics[width=\linewidth]{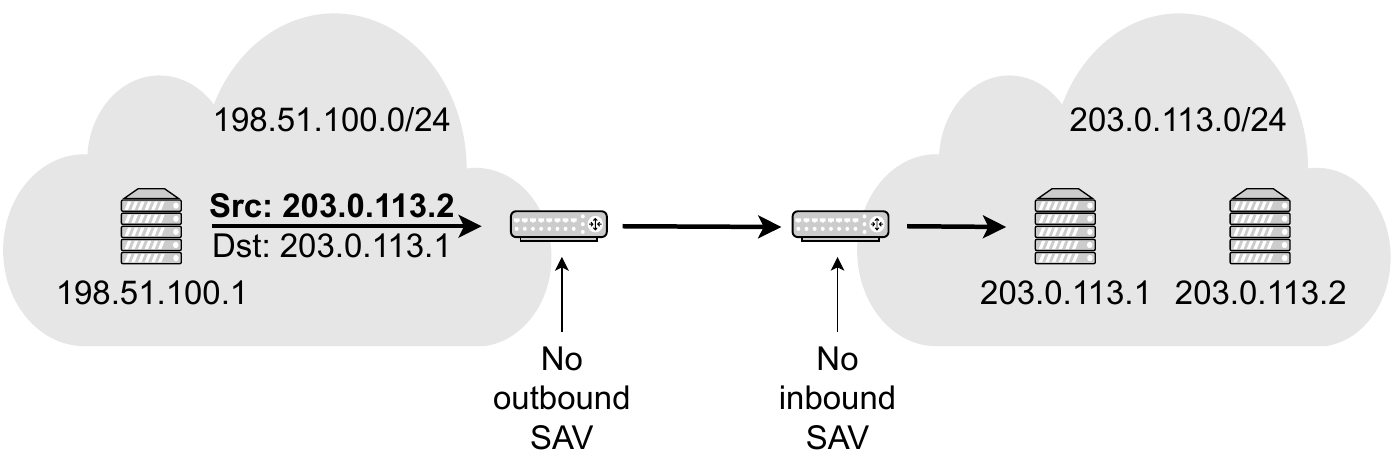}
        \caption{No SAV performed at all: spoofed IP packets can reach the victim server for further exploitation.}
        \label{twonetworks:nosav}
    \end{subfigure}
    
    \begin{subfigure}[b]{0.49\textwidth}
        \includegraphics[width=\linewidth]{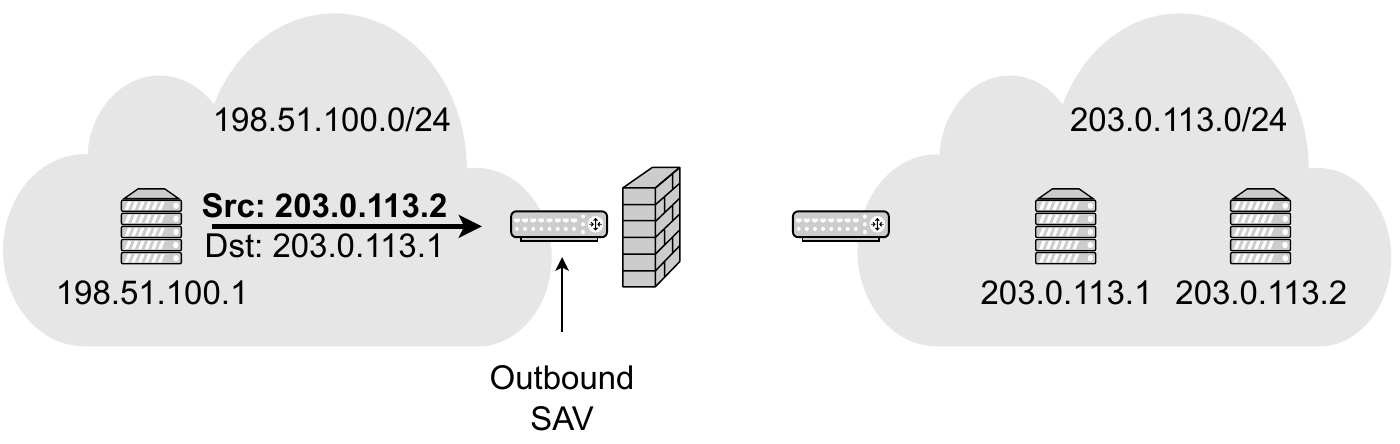}
        \caption{\textit{Outbound} SAV at the edge of the attacker network: it blocks attacks based on IP spoofing. However, for entire elimination, \textit{all} networks would need to implement outbound SAV.}
        \label{twonetworks:outbound}
    \end{subfigure}
    
    \begin{subfigure}[b]{0.49\textwidth}
        \includegraphics[width=\linewidth]{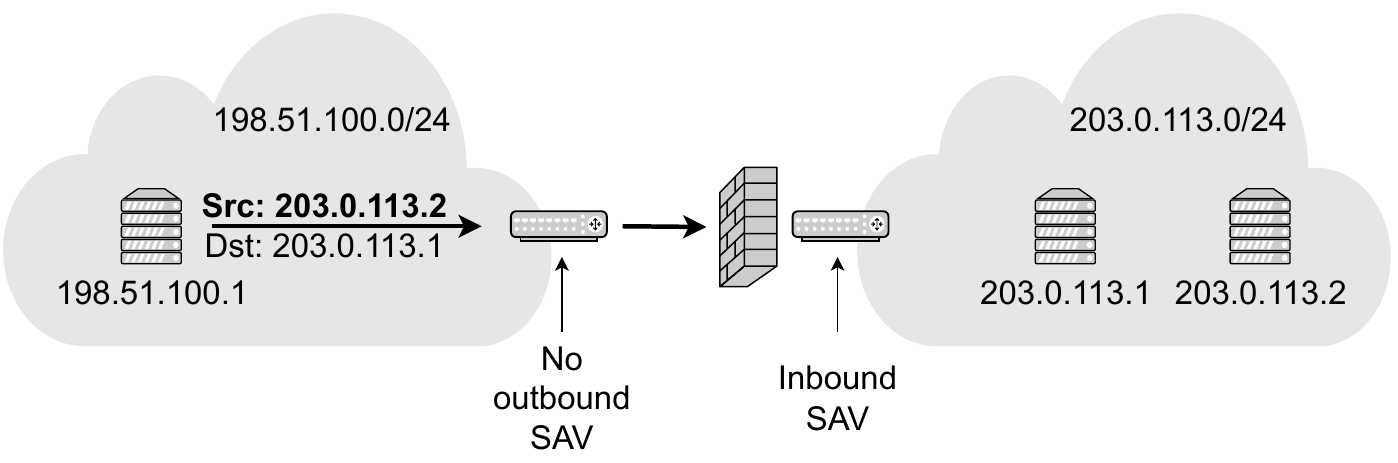}
        \caption{\textit{Inbound} SAV at the edge of the victim network: it directly protected the network against external attackers trying to spoof internal IP addresses.}
        \label{twonetworks:inbound}
    \end{subfigure}
    \caption{Attack scenario using IP spoofing (a) and mitigation strategies using SAV (b,~c). The attacker tries to exploit a victim server \texttt{203.0.113.1} by spoofing the source address of a legitimate client \texttt{203.0.113.2}, hoping to go beyond IP-based access control. Deploying \textit{inbound} SAV at the edge of the victim network (c) is directly effective at protecting servers in this network.
    }
    \label{twonetworks}
\end{figure}

The first prominent DoS attacks started appearing in the 1990s, even though the Internet was still relatively small. These attacks included ICMP echo, SYN, and UDP floods, among others~\cite{2000dos}. As the Internet community was concerned about effective ways to fight such attacks, in 2000 Senie and Ferguson proposed network ingress filtering (documented in the Best Current Practice 38~\cite{Ferguson:2000:NIF:RFC2827}) as a means of mitigating DoS attacks involving source IP address spoofing. The proposed solution was to only allow traffic ``which originates from a downstream network to known, and intentionally advertised, prefix(es).'' \cite{Ferguson:2000:NIF:RFC2827} If deployed universally, it would greatly decrease the effectiveness of spoofing attacks. However, the authors acknowledged that in some cases, such as asymmetric routing, multihoming, or mobile IP, implementing ingress filtering is challenging. In the subsequent Best Current Practice 84~\cite{Baker:2004:IFM:RFC3704}, Baker and Savola presented several ways to implement ingress filtering: static ingress access lists and reverse path forwarding.

Ingress filtering, commonly referred to as Source Address Validation (SAV)~\cite{tracefilter}, can be applied in two directions with respect to a given network: \textit{outbound} (to prevent spoofed packets from leaving the network) and \textit{inbound} (to drop spoofed packets coming from outside the network). These two ways of deploying SAV protect networks from different threats.

First, attackers benefit from the absence of \textit{outbound} SAV to launch amplification and reflection DDoS attacks based on IP spoofing. They randomly choose globally routable addresses as sources and overload target services, effectively making them unable to process genuine requests. Alternatively, they use open services prone to amplification~\cite{hell,Kuhrer:2014:EHR:2671225.2671233} to which they send requests with the source IP address of the victim. Consequently, the victim is overloaded with the traffic coming from reflection/amplification services rather than directly from attackers or botnets controlled by them. In both scenarios, the origin of the attack is not traceable. One of the most successful attacks against Google resulted in 2.5 Tb/s traffic~\cite{googleddos}. Attackers sent spoofed requests with Google IP addresses as sources to 180K CLDAP, DNS, and SNMP servers. 

On the other hand, the absence of \textit{inbound} SAV allows intruders to impersonate internal hosts, which may reveal information about the inner network structure and the presence of closed services, such as DNS resolvers, even though they only accept queries from predefined clients. This is the attack scenario we are interested in, shown in Figure~\ref{twonetworks:nosav}. The attacker, located in the network \texttt{198.51.100.0/24}, sends a packet to the victim \texttt{203.0.113.1} in \texttt{203.0.113.0/24}. The attacker spoofs the source IP address of a legitimate client of the victim server, for instance \texttt{203.0.113.2}, to be able to exploit the server.

The absence of SAV for inbound traffic may have serious security consequences when combined with the DNS Unchained attack \cite{unchained}, the NXDOMAIN attack (also known as the Water Torture Attack)~\cite{LuoWXCYT18}, or NXNSAttack~\cite{NXNSAttack}. These attacks result in Denial-of-Service against both recursive resolvers and authoritative servers. The NXNSAttack exploits the way recursive resolvers deal with referral responses (domain delegations) that provide the mapping between a given domain name and its authoritative nameserver without a glue record, i.e., the IP addresses of the nameserver. The maximum packet DDoS amplification factor of the NXNSAttack attains 1,620~\cite{NXNSAttack}. It also saturates the cache of the resolver, even the closed one, if the attack uses source IP spoofing and inbound SAV is not in place.

The possibility of impersonating a host on the victim network can also assist in the zone poisoning attack~\cite{Korczynski:2016:ZPN:2987443.2987477}. A master DNS server, authoritative for a given domain, may be configured to accept non-secure DNS dynamic updates from a DHCP server on the same network \cite{rfc2136}. Thus, sending a spoofed update from the outside with an IP address of that DHCP server will modify the content of the zone file \cite{Korczynski:2016:ZPN:2987443.2987477}. The attack may lead to domain hijacking. Another way to target closed resolvers is to perform DNS cache poisoning~\cite{kaminsky}. An attacker can send a spoofed DNS request for a specific domain to a closed resolver, followed by forged replies before the arrival of the response from the genuine authoritative server. In this case, the users who query the same domain will be redirected to where the attacker specified until the forged DNS entry reaches its Time To Live~(TTL).

IP spoofing of inbound traffic can be combined with any vulnerable protocol (e.g., NTP, SNMP, SSDP \cite{hell}, FTP, HTTP, Telnet \cite{tcp-hand}, etc.) to launch, among others, self-directed amplification DDoS or attacks against other hosts in the same network. For~example, NTP is known for its high amplification factor up to 4,670. An attacker, sending spoofed requests on behalf of the victim trusted by private NTP servers, can generate massive traffic towards the victim in the same network~\cite{hell,Kuhrer:2014:EHR:2671225.2671233}.

Deploying SAV at the network edge can mitigate these impersonation attacks. As shown in Figure~\ref{twonetworks:outbound}, \textit{outbound} SAV near the source of the traffic filters out forged packets and thus prevents IP spoofing-based attacks. However, all networks in the Internet would need to deploy outbound SAV, a significant operational challenge, which limits SAV deployments by network operators. Lichtblau et al. surveyed 84 network operators to check whether they deployed SAV and learn what challenges they faced~\cite{Lichtblau}. The reasons for not performing packet filtering included incidentally filtering out legitimate traffic, equipment limitations, and the lack of a direct economic benefit---in case of outbound SAV, a compliant network cannot become an attack source, but it can still be attacked itself, which creates few incentives to become compliant. On the other hand, deploying \textit{inbound} SAV at the destination network can directly protect the network itself from impersonation attacks, as shown in Figure~\ref{twonetworks:inbound}. Thus, there is a direct economic incentive to deploy inbound SAV for a network operator.

\section{Related Work \label{sec:related}}

\subsection{Measuring Source Address Validation Deployment \label{sec:relatedSAV}}

Table~\ref{related_work} summarizes several existing methods to infer SAV deployment. They differ in terms of the filtering direction (inbound/outbound), whether they confirm the presence or the absence of SAV, whether measurements can be done remotely or on a vantage point inside the tested network, and if the method relies on existing network misconfigurations.

The Spoofer project~\cite{bb-spoofer-sruti,spoofer_new,Spoofer} deploys a client-server infrastructure mainly based on volunteers (and ``crowdworkers'' hired for one study through five crowdsourcing platforms~\cite{marketplaces}) that run the client software from inside a network. To test outbound SAV compliance, the active probing client sends both unspoofed and spoofed packets to the Spoofer server either periodically or when it detects a new network. The server inspects received packets (if any) and analyzes whether filtering disables spoofing and to what extent~\cite{Beverly:2009:UED:1644893.1644936}. For each client running the software, the Spoofer identifies its /24 IPv4 address block (or /40 for IPv6) and the autonomous system number (ASN). It makes outbound SAV test results publicly available~\cite{Spoofer}. Testing inbound SAV compliance operates in the opposite direction---the Spoofer server sends packets to the client with spoofed source addresses belonging to the client network.  The authors do not make the results public to protect vulnerable networks. This approach identifies the absence and the presence of SAV in both directions. The results obtained by the Spoofer project provide the most confident picture of the deployment of outbound SAV and have covered tests from 9,751 ASes and 218 countries since 2015. However, the network administrators who are not aware of the dangers of spoofing or those who do not deploy SAV are less likely to run Spoofer in their networks, which means that the dataset may be not representative of the whole Internet.

\begin{table}[t]
\caption{Methods to infer  deployment of Source Address Validation} 
\label{related_work} 
\scriptsize
\centering 
\setlength{\tabcolsep}{6pt}
\begin{tabular}{lcccc}
 \toprule
\multirow{3}{5em}{\textbf{Method}} &
\multirow{3}{4em}{\textbf{Direction}} &
\multirow{3}{4em}{\textbf{Presence/ Absence}} &
\multirow{3}{3em}{\textbf{Remote}} &
\multirow{3}{5em}{\textbf{Relies on misconfigurations}} \\
&&&&\\
&&&&\\
 \midrule
 Spoofer~\cite{bb-spoofer-sruti,spoofer_new,Spoofer} & both & both & no & no\\ 
 Forwarder-based~\cite{mauch,Kuhrer:2014:EHR:2671225.2671233} & outbound & absence & yes & yes\\
 Traceroute loops~\cite{loops} & outbound & absence & yes & yes\\
 Passive detection~\cite{Lichtblau} & outbound & absence & no & no\\
 Spoofer-IX~\cite{Lucas} & outbound & absence & no & no \\
 DSAV~\cite{behind} & inbound & absence & yes & no \\
 Our method~\cite{closed}& inbound & both & yes & no\\
 \bottomrule
\end{tabular}
\end{table}

A more practical approach is to perform such measurements remotely. K\"uhrer et al.~\cite{Kuhrer:2014:EHR:2671225.2671233} scanned for open DNS resolvers, as proposed by Mauch~\cite{mauch} to detect the absence of outbound SAV. They leveraged misconfigured forwarding resolvers that forward a request to a recursive resolver with either i) the packet source address not changed to its own address or ii) the response to the client sent with the source IP of the recursive resolver \cite{Kuhrer:2014:EHR:2671225.2671233,saving}. They fingerprinted those forwarders and found out that they were mostly embedded devices and routers. Misconfigured forwarders originated from 2,692 autonomous systems. We refer to this technique as \textit{forwarder-based}.

Lone et al.~\cite{loops} proposed another method that does not require a vantage point inside a tested network. When a packet is sent to a customer network with a routable but not allocated address, it is sent back to the provider router without changing its source IP address. The packet, having the source IP address of the machine that sent it, should be dropped by the router because the source IP does not belong to the customer network. The method detected 703 autonomous systems not deploying outbound SAV.

While the above-mentioned methods rely on actively generated (whether spoofed or not) packets, Lichtblau et al.~\cite{Lichtblau} passively observed and analyzed inter-domain traffic exchanged between more than 700 networks at a large interconnection point (IXP). They classified observed traffic into bogon, unrouted, invalid, and valid based on the source IP addresses and AS paths. The most conservative estimation identified 393 networks that generated invalid traffic. Müller et al.~\cite{Lucas} developed Spoofer-IX, another methodology to detect spoofing at the IXP level. Their traffic classification took into account AS business relationships, asymmetric routing, and traffic engineering. Deployed at one mid-sized IXP during five weeks, it measured 40 Mb/s as the upper bound of spoofed traffic.

In the concurrent work, Deccio~et al.~\cite{behind} remotely measured  the absence of inbound SAV. They issued DNS requests towards 11 million IPv4 and 785K IPv6 addresses and classified 26K IPv4 (3.9K IPv6) autonomous systems as vulnerable to spoofing of inbound traffic. The authors refer to inbound SAV as  ``destination-side source address validation'' or DSAV for short.

We are the first to propose a remote method (no vantage points needed in the tested networks) to estimate the deployment of inbound SAV that does not rely on existing misconfigurations. Instead, we take advantage of local DNS resolvers in remote networks (both open and closed) to infer the absence or the \textit{presence} of SAV either in transit networks or at the destination network edge. Our measurements cover the whole routable IPv4 address space and more than 270M responsive IPv6 addresses.

\subsection{Identifying Dual-Stack Servers}

To compare the SAV deployment status for IPv4 and IPv6, we identify seemingly dual-stack DNS  resolvers.

Several researchers used DNS to obtain candidate (IPv4, IPv6) address pairs that likely indicate to be the same physical machine (also called dual-stack). Berger et al.~\cite{nameserver} developed two passive and active techniques to find such pairs. They deployed the passive method over the existing production infrastructure consisting of a two-level authoritative nameserver hierarchy in which the first-level server, reachable over IPv4, returns records of the second-level server. In its DNS response, it also encodes the IPv4 address of the contacting client. Each request arriving at the second-level nameserver over IPv6 reveals the initial IPv4 query. This method is not restricted to open resolvers and does not actively generate additional DNS requests. The method discovered 674K candidate pairs during a period of six months. The second active technique relies on sending requests to open resolvers for such multi-level domains, which implies switching between IPv4 and IPv6 protocols using \texttt{CNAME} records. In a one-day measurement session, they probed 200 times 7K open resolvers and revealed 41K address pairs.

Hendriks et al.~\cite{ipv6_or} enumerated the population of open IPv6 resolvers to analyze whether they could be used as efficient DDoS amplifiers. They first performed an Internet-wide scan to find open resolvers over IPv4 and queried them for specifically-crafted domains that could only be reached by traversing from IPv4 to IPv6. This method discovered 1.49M unique candidate pairs and 1,038 unique IPv6 resolvers.

The two approaches described above do not necessarily find dual-stack machines (also called siblings) but rather dual-stack candidate pairs. There is a need to validate those results. Beverly et al.~\cite{server_siblings} proposed a technique that is not limited to DNS resolvers but also relies on collected TCP-level information such as option signatures and timestamps. The algorithm was 97\% accurate in identifying sibling relationships. In 2017, Scheitle et al.~\cite{quirin} developed a machine-learning algorithm that also gathered various TCP-level features (options, timestamp clock frequency, timestamp value, clock offset, etc.) and calculated a variable clock skew. The precision of the algorithm exceeded~99\%.

Czyz et al.~\cite{back_door} showed that the IPv6 Internet is more open than IPv4. They developed two candidate lists: router IP pairs and pairs derived from DNS zone files. They probed all addresses on various ports for services expected to run on routers and DNS servers. To ascertain that some pairs were indeed dual-stack machines, they collected fingerprinting information of the following applications: HTTP, HTTPS, SNMP, NTP, SSH, and MySQL. Based on this information, they confirmed that 96\% of router and 97\% of nameserver pairs, open on at least one of the ports, were the same physical machines.

To compare the SAV deployment status for IPv4 and IPv6, we have deployed a two-level hierarchical DNS zone infrastructure that forces a recursive resolver to switch from IPv4 to IPv6 (and vice versa) to resolve our domain names. Whenever we detect that an IPv4 or IPv6 resolver is also reachable over IPv6 and IPv4, respectively, we consider such address pairs to be dual-stack candidates. We send spoofed and non-spoofed packets to target both open and closed resolvers. We then fingerprint them on different ports to gather evidence on whether each pair belongs to the same physical machine.

\section{Methodology}\label{sec:methodology}

In this section, we present the methodology for identifying the networks that deploy (or not) inbound SAV, locating dual-stack DNS resolvers, and fingerprinting them.

\subsection{IPv4 Spoofing Scan \label{sec:spoof_scan}}

The core idea of the spoofing scan is to send hand-crafted DNS \texttt{A} requests with spoofed source addresses to all the routable hosts. We have developed an efficient scanner capable of sending spoofed DNS requests in bulk. It runs on a machine in a network that does not deploy outbound SAV so that we can send packets with spoofed IP addresses. We make the scanner available to the interested researchers upon request. When our query reaches a recursive DNS resolver inside a network not deploying inbound SAV, the resolver starts the resolution process. We observe the query on our authoritative DNS servers. To prevent caching and to identify the true originator in case of forwarding, we always query a unique domain name composed of: a random string, a hex-encoded resolver IP address (the destination of our query), a scan identifier, an IP version subdomain and a domain name itself. The encoded IP address lets us identify forwarders: if the IP address seen on our authoritative nameservers is not the same as originally queried (extracted from the domain name), we know that the query destination is a forwarder. An example domain name is \texttt{\small{dklL56.cb007101.s1.v4.drakkardnsv4.com}}.

\begin{figure}[!t]
\centering
\includegraphics[width=\linewidth]{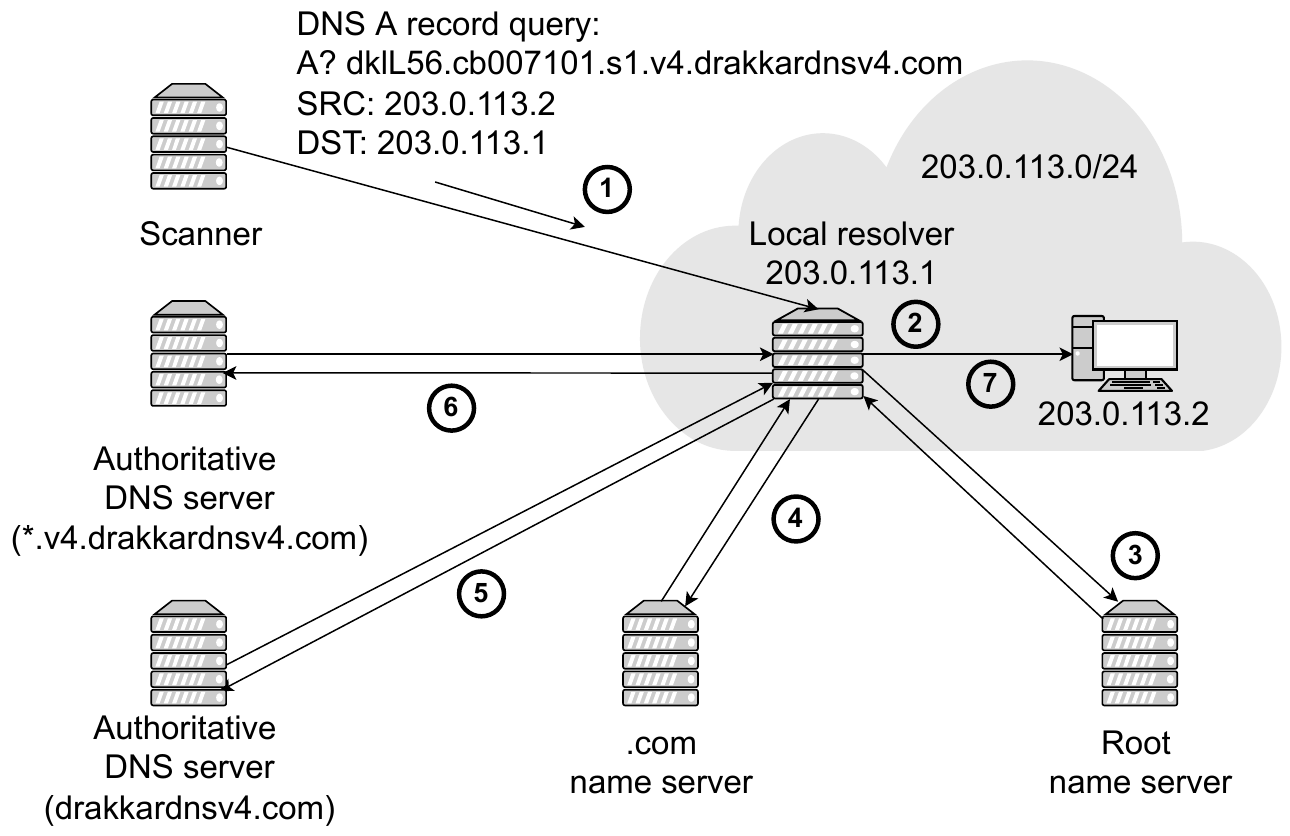}
\caption{Setup of the spoofing scan over IPv4. We set up devices on the left-hand side (scanner, authoritative nameservers) and do not have control over the remaining infrastructure.}
\label{setup}
\end{figure}

Figure~\ref{setup} shows the scanning setup for the example \texttt{\small{203.0.113.0/24}} network. In Step~1, the scanner sends one spoofed packet to each potential host of the network (256 packets in total). The spoofed source IP address is always the next one after the destination. If we reach the last IP address of the network (e.g. \texttt{\small{203.0.113.255}} for \texttt{\small{203.0.113.0/24}}), we go back to the beginning and use \texttt{\small{203.0.113.1}} as a spoofed source IP. When the scanner sends the spoofed packet containing the DNS query, there are four possible cases:

\textit{1) Packet filtering in transit network or random losses:} The spoofed packet can be filtered anywhere in transit or dropped due to reasons not related to IP spoofing such as network congestion~\cite{Beverly:2009:UED:1644893.1644936}.

\textit{2) Inbound SAV at the network edge:} When the spoofed DNS packet arrives at the destination network edge (therefore, it has not been filtered anywhere in transit), the packet filter inspects the packet source address and detects that such a packet cannot arrive from the outside because the address block is allocated inside the network. Thus, the filter drops the packet.

\textit{3) No inbound SAV at the network edge and no DNS resolver inside the network:} The packet enters the network, but there is no local DNS resolver in the tested network, so the DNS query is not resolved. In some cases, the DNS resolver is present but may be configured to refuse queries coming from its local area network (for example, if the whole separate network is dedicated to the infrastructure), so the packet is also dropped.

\textit{4) No inbound SAV at the network edge and the destination host is a DNS resolver:} The scanner eventually reaches all the hosts in the network and the local DNS resolver, if there is one (\texttt{\small{203.0.113.1}} in Figure~\ref{setup}). When the local resolver receives a DNS \texttt{A} record request (Step~2) from a host on the same network (\texttt{\small{203.0.113.2}}), it performs query resolution (Steps~3--6) that eventually reaches our authoritative nameserver. The local resolver sends the response back to the source address (Step~7). 

Note that only the last case allows \textit{inferring the absence of inbound SAV} and we cannot distinguish between the first three cases.

There are two types of resolvers: forwarders that forward queries to other recursive resolvers and non-forwarders that directly resolve queries they receive. The DNS resolver (\texttt{\small{203.0.113.1}}) in Figure~\ref{setup}  is a non-forwarder. It inspects the query that looks as if it were sent from \texttt{\small{203.0.113.2}} and performs the resolution by iteratively querying the root (Step~3) and the top-level domain (Step~4) servers until it reaches our authoritative nameservers in Steps~5 and~6. Alternatively, if \texttt{\small{203.0.113.1}} were a forwarder, it would forward the query to another recursive resolver that would repeat the same procedure as described above for non-forwarders. In Step~7, the DNS \texttt{A} query response is sent to the spoofed source (\texttt{\small{203.0.113.2}}).

Our goal is to scan the whole IPv4 address space, yet taking into account only globally routable and allocated address ranges. We use the data provided by the RouteViews project~\cite{routeviews} to get all the IPv4 blocks currently present in the BGP routing table and send spoofed DNS requests to all the hosts in these prefixes.

\begin{figure}[!t]
    \centering
    \includegraphics[width=\linewidth]{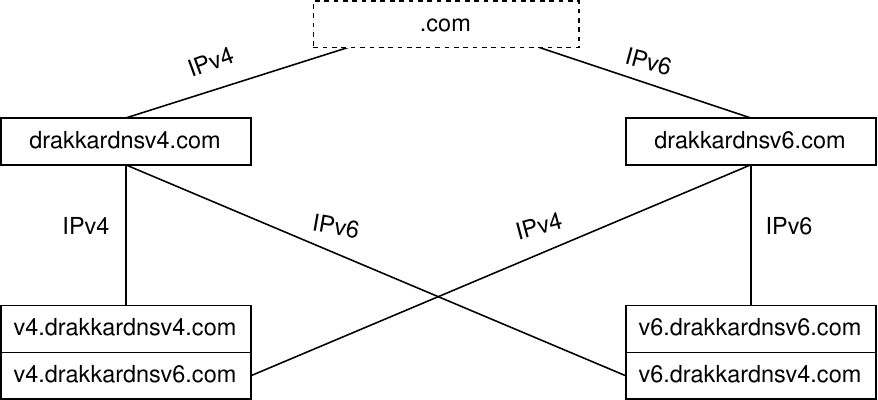}
    \caption{DNS zone setup. Rectangles with solid lines represent authoritative nameservers for the corresponding DNS zones (domain names) under our control. The \texttt{.com} zone (dashed) only contains glue records (IP addresses) of nameservers authoritative for our domains and is out of our control. Edges indicate the network protocol (IPv4 or IPv6) needed to reach a given zone.}
    \label{zones}
\end{figure}

\subsection{IPv6 Spoofing Scan \label{sec:spoof_v6}}

The complete scan of the IPv6 space is not possible, even considering only the networks present in the BGP routing table. We scan two sets of IPv6 addresses: those discovered by traversing from IPv4 to IPv6 (described in Section~\ref{sec:dsscan}) and the addresses from the IPv6 Hitlist Service~\cite{hitlist}. On  the  day  of the measurement, the IPv6 Hitlist Service contained 270M addresses for scanning.

We send spoofed DNS \texttt{A} requests to all hosts from the two to-scan datasets and spoof the source to be the next IP address after the target. The format of the domain name is similar to the IPv4 one: \texttt{\small{qGPDBe.long\_int(ipv6).s1.v6. drakkardnsv6.com}}. We represent the IPv6 address as a long integer to identify the initial query destination uniquely and to distinguish forwarders from non-forwarders. We still send requests for the DNS \texttt{A} record, as changing the network protocol does not influence the retrieved resource records.

\subsection{Open Resolver Scan \label{sec:orscan}}

In parallel to the spoofing scan, we perform an open resolver scan over IPv4 and IPv6 by sending DNS \texttt{A} requests with genuine source IP addresses of the scanner. To avoid temporal changes, we send a non-spoofed query just after the spoofed one to the same host. The format of a non-spoofed query is almost the same as the spoofed one, the only difference is the scan identifier (\texttt{\small{n1}} referring to a non-spoofed scan identifier instead of \texttt{\small{s1}}). Example domain names are  \texttt{\small{qGPDBe.cb007101.n1.v4.drakkardnsv4.com}} for IPv4 and  \texttt{\small{qGPDBe.long\_int(ipv6).n1.v6.drakkardnsv6.com}} for IPv6. If we receive a non-spoofed request on our authoritative nameservers, it means that we have reached an open resolver. Moreover, if this open resolver did not resolve the spoofed query, we infer the presence of inbound SAV either in transit or at the tested network edge.

\begin{figure}[!t]
    \centering
    \includegraphics[width=\linewidth]{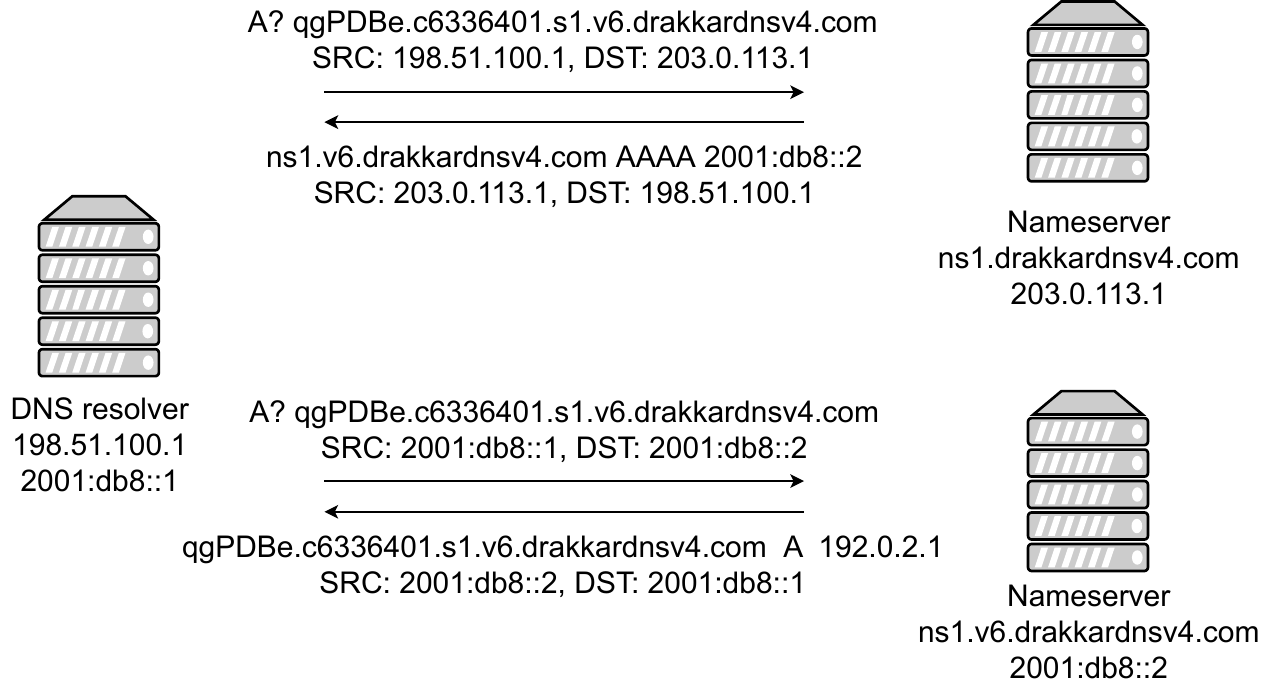}
    \caption{Domain name resolution that requires switching from IPv4 to IPv6. The DNS resolver on the left-hand side contacts the \texttt{ns1.drakkardnsv4.com} nameserver over IPv4. It does not receive the answer to the \texttt{A} request directly, but rather a referral to the \texttt{ns1.v6.drakkardnsv4.com} nameserver only reachable over IPv6.}
    \label{dual-stack-fig}
\end{figure}

\begin{table*}[t]
    \caption{Types of discovered DNS resolvers} 
    \label{resolvers} 
    \scriptsize
    \centering 
    \begin{tabular}{lcccccc}
         \toprule
         & \multirow{2}{*}{\textbf{\# scanned hosts }}  &
         \multirow{2}{*}{\textbf{Total DNS resolvers}} & \textbf{Closed resolvers in networks}  &
         \textbf{Open resolvers in networks} & \textbf{Open resolvers in networks} \\
         &&& \textbf{without inbound SAV} & \textbf{without inbound SAV} & \textbf{with inbound SAV}\\
         \midrule
         IPv4 & 2,831,160,434 & 7,871,673 & 2,522,869 & 3,970,827 & 1,377,977 \\
         \midrule
         IPv6 & 270,703,379 & 115,610 & 99,718 & 8,977 & 6,915\\
         \bottomrule
    \end{tabular}
\end{table*}

\subsection{Identifying Dual-Stack Candidates \label{sec:dsscan}}

To compare the level of SAV deployment for IPv4 and IPv6 at the machine level, we collect (IPv4, IPv6) address pairs likely belonging to the same physical machines. We do so by deploying two-level DNS zones as shown in Figure~\ref{zones}. 

We set up two parent domains (\texttt{\small{drakkardnsv4.com}} and \texttt{\small{drakkardnsv6.com}}) on two distinct servers. Each domain has only one glue record (i.e., IP address of the nameserver) configured via the registrar control panel: an IPv4 address for \texttt{\small{drakkardnsv4.com}} and an IPv6 address for \texttt{\small{drakkardnsv6.com}}. For example, the authoritative nameserver of \texttt{\small{drakkardnsv4.com}} is \texttt{\small{ns1.drakkardnsv4.com}}. Likewise, the authoritative nameserver of \texttt{\small{drakkardnsv6.com}} is \texttt{\small{ns1.drakkardnsv6.com}}. Thus, at the DNS level, each nameserver can only be reached over one network layer protocol (IPv4 or IPv6) but not over both. Two more servers host child DNS zones: \texttt{\small{v4.drakkardnsv4.com}} and \texttt{\small{v6.drakkardnsv6.com}}. These two domains, each reachable over only IPv4 and IPv6, respectively, are used for IPv4 and IPv6 scans. We then add two more child domains that require traversal from IPv4 to IPv6 and the other way round: \texttt{\small{v4.drakkardnsv6.com}} and \texttt{\small{v6.drakkardnsv4.com}}. The \texttt{A} record of   \texttt{\small{ns1.v4.drakkardnsv6.com}} is added to the \texttt{\small{drakkardnsv6.com}} parent zone, while the \texttt{AAAA} record of \texttt{\small{ns1.v6.drakkardnsv4.com}} is added to the \texttt{\small{drakkardnsv4.com}} zone.

Figure~\ref{dual-stack-fig} shows how a dual-stack recursive resolver on the left (configured with IPv4 and IPv6 addresses)  resolves our \texttt{\small{qgPDBe.c6336401.s1.v6.drakkardnsv4.com}} domain name. We assume that it previously obtained the IPv4 address of  \texttt{\small{ns1.drakkardnsv4.com}}. The resolver contacts this nameserver over IPv4 asking for the \texttt{\small{A}} record of the queried domain name. The nameserver cannot directly provide the answer. Instead, it refers to \texttt{\small{ns1.v6.drakkardnsv4.com}}, only configured with an IPv6 glue record. The resolver, if capable of doing so, now has to switch to IPv6 and contact \texttt{\small{ns1.v6.drakkardnsv4.com}} at \texttt{2001:db8::2}. If successful, the resolver receives the final response at its IPv6 address (\texttt{\small{2001:db8::1}}).

During IPv4 and IPv6 spoofing scans, we continuously  analyze traffic captures from our nameservers to avoid IP address churn~\cite{Kuhrer:wild} and identify DNS resolvers that processed our requests. We only use non-forwarders as possible dual-stack candidates as they are less likely to be a part of a complex DNS infrastructure, not visible from our authoritative nameservers, which includes, but is not limited to, load balancing and DNS cache sharing~\cite{back_door}. As soon as we detect non-forwarders, we send them requests with domains that imply switching to the other IP version. Examples of such domains are \texttt{\small{qgPDBe.long\_int(ipv6).nf.s1.v4.drakkardnsv6.com}} and \texttt{\small{qgPDBe.c6336401.nf.s1.v6.drakkardnsv4.com}}.We later retrieve source IP addresses that resolved these domains. Together with the addresses encoded in domain names, they form (IPv4, IPv6) candidate pairs. To further enrich the IPv6 Hitlist, we request all the identified IPv4 resolvers (whether forwarders or non-forwarders) to resolve a subdomain of \texttt{\small{v6.drakkardnsv4.com}}.

\subsection{Fingerprinting\label{sec:fingerprint}}

Some services expose banners and other information such as software versions, operating systems they run on, public keys, certificates, etc. We perform a preliminary measurement campaign and scan 1K (IPv4, IPv6) candidate pairs for most common ports using \texttt{nmap} \footnote{\url{https://nmap.org/book/nmap-services.html}}.
Some of the open ports are 22 (SSH), 53 (DNS), 80 (HTTP), 443 (HTTPS), and 587 (SMTP). We consider the fraction of the remaining open ports negligible and not suitable for fingerprinting. We additionally scan for port 123 (NTP), as NTP is a protocol commonly used to amplify DDoS attacks~\cite{hell}\cite{Kuhrer:2014:EHR:2671225.2671233}, and find that more than 10\% of addresses had port 123 open. We explain below how we gather fingerprinting information.

\textit{1) DNS:} Two types of DNS queries can be used to identify remote DNS servers. The first one is a reverse DNS lookup that can determine the domain name associated with an IP address. It requires querying for a pointer (\texttt{PTR}) resource record. It is a recommended practice to have a hostname configured for every IP address~\cite{common_dns} and it was shown that 1.2B responsive IPv4 addresses (28.17\% of the whole IPv4 space) have an associated \texttt{PTR} record~\cite{rdns}. We perform reverse DNS lookups for each (IPv4, IPv6) sibling candidate pair and check for an exact match between returned domain names. It is common for shared hostnames to represent a single machine~\cite{back_door}. The second fingerprinting query is a \texttt{CHAOS} class \texttt{TXT} record for \texttt{version.bind} name. It is one of the specific queries originally introduced as a debugging tool for network administrators~\cite{RFC4892}. Contrary to what its name suggests, it is implemented in different DNS software, not only in BIND. Unless explicitly hidden, a DNS resolver replies with the exact installed software version. The example return values include ``9.11.10-RedHat-9.11.10-1.fc29'' or ``unbound 1.10.0''. We look for candidate pairs in which the same version is displayed for both. We ignore the cases when an arbitrary string is returned.

\textit{2) NTP:} We fingerprint servers over UDP port 123 using \texttt{nmap}. The NTP standard~\cite{ntp} specifies the \texttt{version} packet header variable as a ``3-bit integer representing the NTP version number.'' We use the \texttt{ntp-info} script \footnote{\url{https://nmap.org/nsedoc/scripts/ntp-info.html}} 
to collect the version of the running NTP deamon, as well as system and processor types.

\begin{table*}[t]
    \caption{Deployment of inbound SAV}
    \label{inconsistencies} 
    \scriptsize
    \centering 
    \setlength{\tabcolsep}{8pt}
    \begin{tabular}{lccccccccc}
         \toprule
         \multirow{3}{*}{\textbf{Network Type}} & \multicolumn{2}{c}{\textbf{Consistent absence of}} &  \multicolumn{2}{c}{\textbf{Partial absence of}} & \multicolumn{2}{c}{\textbf{Consistent presence of}} & \multicolumn{2}{c}{\multirow{2}{*}{\textbf{No data}}} & \multirow{3}{*}{\textbf{Total}}\\
         
         & \multicolumn{2}{c}{\textbf{ inbound SAV}} & \multicolumn{2}{c}{\textbf{ inbound SAV}} & \multicolumn{2}{c}{\textbf{inbound SAV}} &\\
         
         \cmidrule(lr){2-3}
        \cmidrule(lr){4-5}
        \cmidrule(lr){6-7}
        \cmidrule(lr){8-9}
         & \textbf{Count} & \textbf{Ratio (\%)} & \textbf{Count} & \textbf{Ratio (\%)} & \textbf{Count} & \textbf{Ratio (\%)} & \textbf{Count} & \textbf{Ratio (\%)} &\\
         \midrule
         IPv4 Autonomous Systems & 21,314 & 31.8 & 11,441 & 17.1 & 2,092 & 3.1 & 32,131 & 48.0 & 66,978\\
         IPv4 BGP prefixes & 152,316 & 17.9 & 45,292 & 5.4 & 39,341 & 4.7 & 609,839 & 72.0 & 846,788\\
         IPv4 /24 networks & 765,233 & 6.9 & 173,239 & 1.5 & 266,498 & 2.4 & 9,948,051 & 89.2 & 11,153,021\\
         \midrule
         IPv6 Autonomous Systems & 4,639 & 24.8 & 127 & 0.7 & 138 & 0.7 & 13,806 & 73.8 & 18,710\\
         IPv6 BGP prefixes & 6,731 & 8.0 & 142 & 0.2 & 274 & 0.3 & 76,526 & 91.5 & 83,673\\
         IPv6 /40 networks & 7,562 & 0.02 & 136 & 0.0002 & 2,874 & 0.006 & 49,408,039 & 99.9 & 49,418,611\\
         \bottomrule
    \end{tabular}
\end{table*}

\textit{3) SMTP:} Port 587 is used for email submission by email clients and servers~\cite{mail}. An extension to SMTP allows secure communication over the Transport Layer Security (TLS) protocol~\cite{smtp_tls}. We use \texttt{openssl} \footnote{\url{https://www.openssl.org}}
to initiate a connection and obtain the server certificate.

\textit{4) HTTP:} We use the ZGrab 2.0 \footnote{\url{https://github.com/zmap/zgrab2}}
application-layer scanner to get home pages, headers, and certificates for the three remaining protocols (HTTP, HTTPS, and SSH). The software initiates a GET request to the potential web server over HTTP. In case of a successful connection, we look for an HTTP \texttt{Server} header field with the running software version.

\textit{5) HTTPS:} Web servers delivering content over the TLS protocol provide more information about the machine in addition to what we can learn with HTTP. The TLS specification~\cite{tls} defines a handshake protocol between the client and the server. The server responds to the client request with the \texttt{\small{ServerHello}} message \cite{infocom}. We retrieve  \texttt{\small{cipher\_suite}} and \texttt{\small{server\_version}} (the TLS version chosen by the webserver based on what is proposed by the client) parameters. We also check the \texttt\small{{Certificate}} message for the returned certificate and \texttt{\small{ServerKeyExchange}} message for the \texttt{\small{tls\_version}}~\cite{back_door} actually used.

\textit{6) SSH:} To establish a connection over port 22, the client and the server must perform a handshake that includes the information on the protocol version~\cite{rfc4253}, which implies sending and receiving identification strings that we retrieve. If the connection is successful, we further collect the server public key fingerprint and the key length~\cite{back_door}.

\section{Inferring the Presence and the Absence of SAV}\label{sec:results}

We have been performing spoofing and open resolver scans since July 2019. For this study, we use data from the scan carried out in March 2020, using the methodology described in Section \ref{sec:methodology}.

\subsection{IPv4 Scan} \label{sec:res_v4}

We have sent two DNS requests (one spoofed and one non-spoofed) to more than 2.8B hosts, excluding roughly 24M addresses from the BGP table as a result of not-to-scan requests from network administrators (see Section~\ref{ethics}). Our \texttt{\small{ns1.v4.drakkardnsv4.com}} authoritative nameserver has received and processed 10.9M spoofed and 9.2M non-spoofed \texttt{A} requests. We define each request as a (source IP address, domain name) tuple. Due to proactive caching or premature querying~\cite{revealed}, DNS resolvers may issue repeating lookups shortly before the TTLs of cached \texttt{A} records expire. Thus, we further analyze only unique requests: 8.7M spoofed and 7.5M non-spoofed.

As each domain name contains the hexadecimally encoded IP address of the query target, we know which DNS resolvers received our requests and processed them. We extract this information from domain names and summarize the number of found DNS resolvers in Table~\ref{resolvers}. In total, we identify 7.9M unique DNS resolvers: 6.5M (2.5M closed and 3.9M open) in networks without inbound SAV and 1.3M open resolvers in networks with inbound SAV in place.

\subsection{IPv6 Scan} \label{sec:res_v6}

During the IPv6 scan, performed immediately after the IPv4 measurement, we have probed 270M IPv6 addresses from the Hitlist Service and 105K addresses learned by traversing from IPv4 to IPv6-only zones as discussed in Section~\ref{sec:dsscan}. We  analyze  all  the (non)-spoofed \texttt{A} requests received  on  \texttt{\small{ns1.v6.drakkardnsv6.com}}. Our authoritative nameserver processed 290K spoofed and 40K non-spoofed \texttt{A} requests. After filtering out duplicates, we get 120K and 23K unique requests, respectively.

For the total of 115K located resolvers, 62K were discovered by traversing from IPv4 to IPv6, 76K from the IPv6 Hitlist Service, and 22K appeared in both groups. The results highlight the added value of the method to identify IPv6 addresses by sending spoofed requests to dual-stack resolvers as explained in Section~\ref{sec:dsscan}. Table~\ref{resolvers} presents the number of resolvers by type. Contrary to results in the IPv4 address space, the great majority of them are closed (100K) and would not be detectable without the proposed spoofing discovery technique. Open resolvers are far less numerous, yet located mostly in networks without inbound SAV in place.

\subsection{Deployment of Inbound SAV\label{sav_depl}}

We associate each resolver IP address with the corresponding /24 IPv4 (/40 IPv6) network, BGP routing prefix, and the autonomous system number using \texttt{pyasn} \footnote{\url{https://github.com/hadiasghari/pyasn}}.
Note that multiple resolvers may belong to a single network/prefix/AS. We define three types of networks/prefixes/ASes with respect to the deployment of inbound SAV. They can be characterized~by:

\textit{1) Consistent absence of inbound SAV:} All the discovered DNS resolvers inside a single network/prefix/AS indicate the absence of inbound SAV.

\textit{2) Partial absence of inbound SAV:} Some resolvers indicate the absence while the others indicate the presence of inbound SAV.

\textit{3) Consistent presence of inbound SAV:} All the discovered DNS resolvers indicate the presence of inbound SAV at the edge of the network under measurement or filtering in transit.

With the proposed method, we cannot unambiguously ascertain whether an entire network/prefix/AS is vulnerable to inbound IP spoofing. However, when reporting the deployment of inbound SAV, we refer to the results of our measurements, i.e., whether they consistently or partially indicate the absence or presence of inbound SAV.

Table~\ref{inconsistencies} presents the scan results classified according to the status of inbound SAV deployment for the IPv4 (first three rows) and IPv6 (last three rows) address spaces. The ``Total'' column indicates the number of routable networks as of February 2020 and ``No data'' corresponds to the networks that did not respond to any query sent by us. The ratios in each of the four groups (consistent absence of inbound SAV, partial absence of inbound SAV, consistent presence of inbound SAV, and no data) are computed based on the ``Total" column. For IPv4, our measurements covered the whole routable address space and we obtained the responses from the majority of autonomous systems (52\%). The coverage of the IPv6 address space is smaller as we scanned the target list of addresses. Nevertheless, a significant ratio of autonomous systems (26.2\%) resolved our queries.

Our measurements indicate that few networks consistently implement inbound SAV and are thus protected from spoofing attacks (see column ``Consistent presence of inbound  SAV'' but note that it includes the cases of filtering in transit). On the contrary, most networks that responded to our requests show the consistent or partial absence of inbound SAV (see columns ``Consistent absence of inbound SAV'' and ``Partial absence of inbound SAV''), whether in the IPv4 or IPv6 address spaces. Overall, 48.9\% IPv4 and 25.5\% IPv6 autonomous systems worldwide are consistently or partially vulnerable to spoofing of inbound traffic. 

Our measurements set a lower bound on the number of networks without inbound SAV for at least three reasons. First, some networks may be vulnerable but they do not contain DNS resolvers. Our queries reach the intended targets but they do not process them according to the DNS specification. Second, some networks may contain closed DNS resolvers, but they only accept requests from other hosts than their own local network. As a result, even if the network is vulnerable and contains a DNS resolver, it will never process the spoofed query. Finally, due to filtering in transit, our spoofed packets never reach networks without SAV with DNS resolvers that would otherwise process our requests. Filtering in transit is also the reason why the number of networks in column ``Consistent presence of inbound SAV'' is an upper bound. As the reported numbers include the cases of filtering in transit, the number of networks deploying inbound SAV at the network edge is actually lower. Some of the cases in ``No data'' column of Table~\ref{inconsistencies} are justified by the three discussed reasons. If we presume a uniform distribution of our measurements, by extrapolating these numbers for the entire IP address space, we obtain over 94\% of IPv4 ASes and 97\% of IPv6 ASes with consistent or partial absence of inbound SAV.

\subsection{Comparison with the Spoofer Project\label{comparisonspoofer}}

We compare the results of our active measurements with the inbound SAV compliance tests performed by the Spoofer project. The Spoofer client-server system provides the most reliable method to infer SAV deployment, as it counts on the presence of vantage points inside tested networks. The server-side Spoofer software sends packets with forged source IP addresses belonging to the IP range of the networks under test. Inbound test results are not publicly available, but we contacted CAIDA and gained access to test data from February-March 2020. In total, 36,073 individual tests were performed. We aggregate the data and keep the latest tests per /24 IPv4 and /40 IPv6 networks. In addition, we only analyze those cases for which the software succeeded in determining the presence or absence of inbound SAV. We found 169 /24 IPv4 and 83 /40 IPv6 networks in common. Both methods agree on the status of inbound SAV deployment for the great majority (83\% for IPv4 and 81\% for IPv6) of those networks. Note that the two methods exhibit certain limitations, such as filtering in transit or packet losses, so, the results may differ.

\begin{figure}[t]
  \centering
  \includegraphics[width=\linewidth]{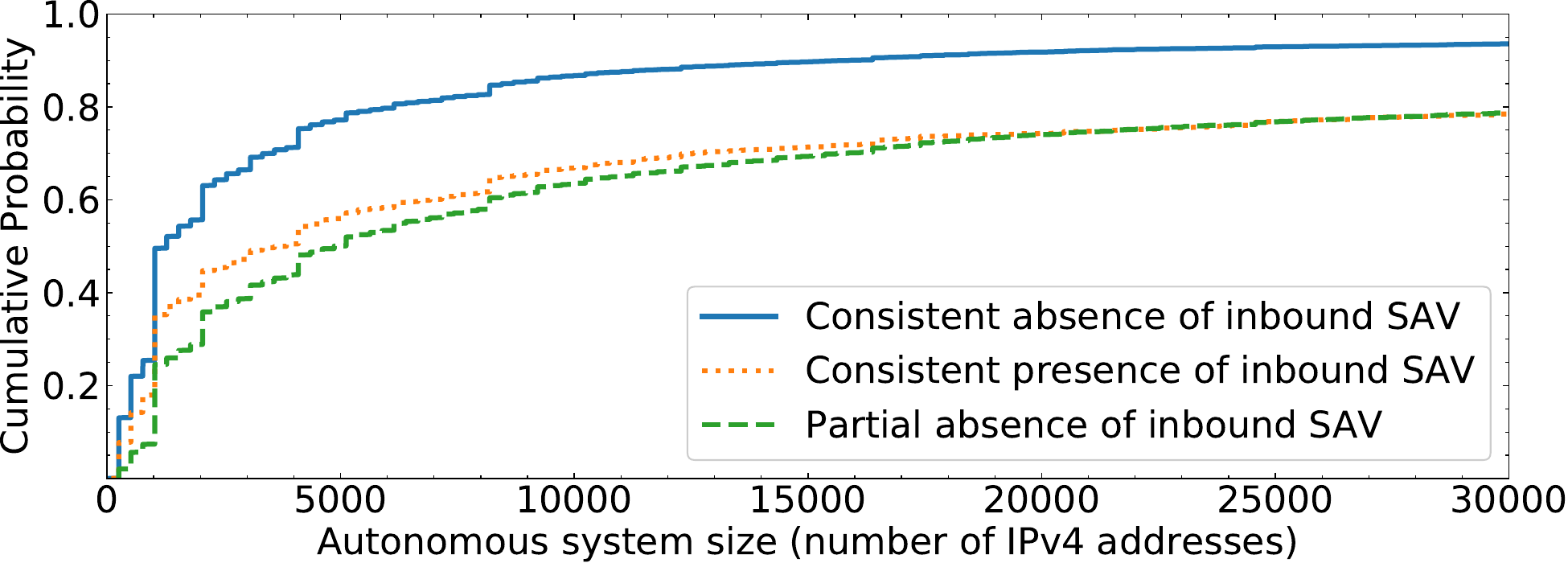}
  \caption{Sizes of IPv4 ASes computed based on the number of unique IPv4 addresses present in the BGP routing table. The cumulative probability shows that ASes with consistent absence of inbound SAV tend to be smaller than other ASes.
  }
  \label{cdf_as_sizes}
\end{figure}

\begin{figure}[t]
  \centering
   \includegraphics[width=\linewidth]{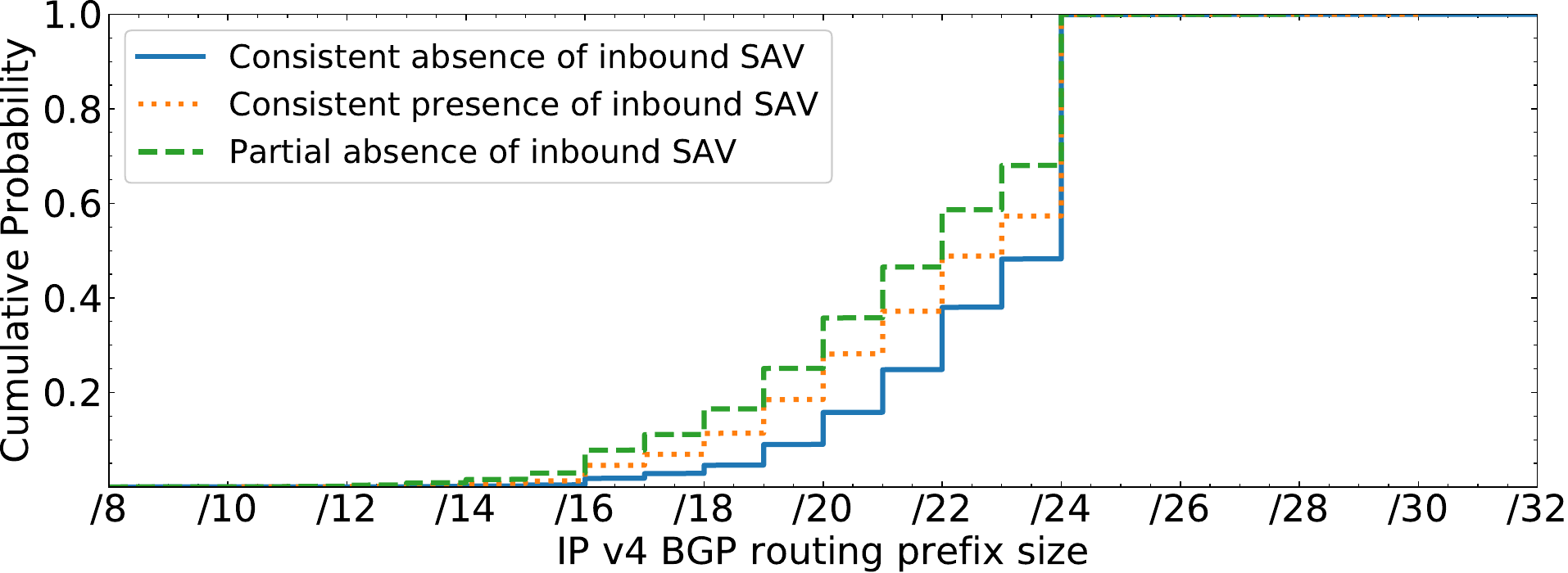}
  \caption{Sizes of the IPv4 longest matching prefixes from the BGP routing table. Larger prefixes are more likely to suffer from partial absence of inbound SAV.
  }
  \label{cdf_prefix_sizes}
\end{figure}

\begin{figure}[t]
  \centering
   \includegraphics[width=\linewidth]{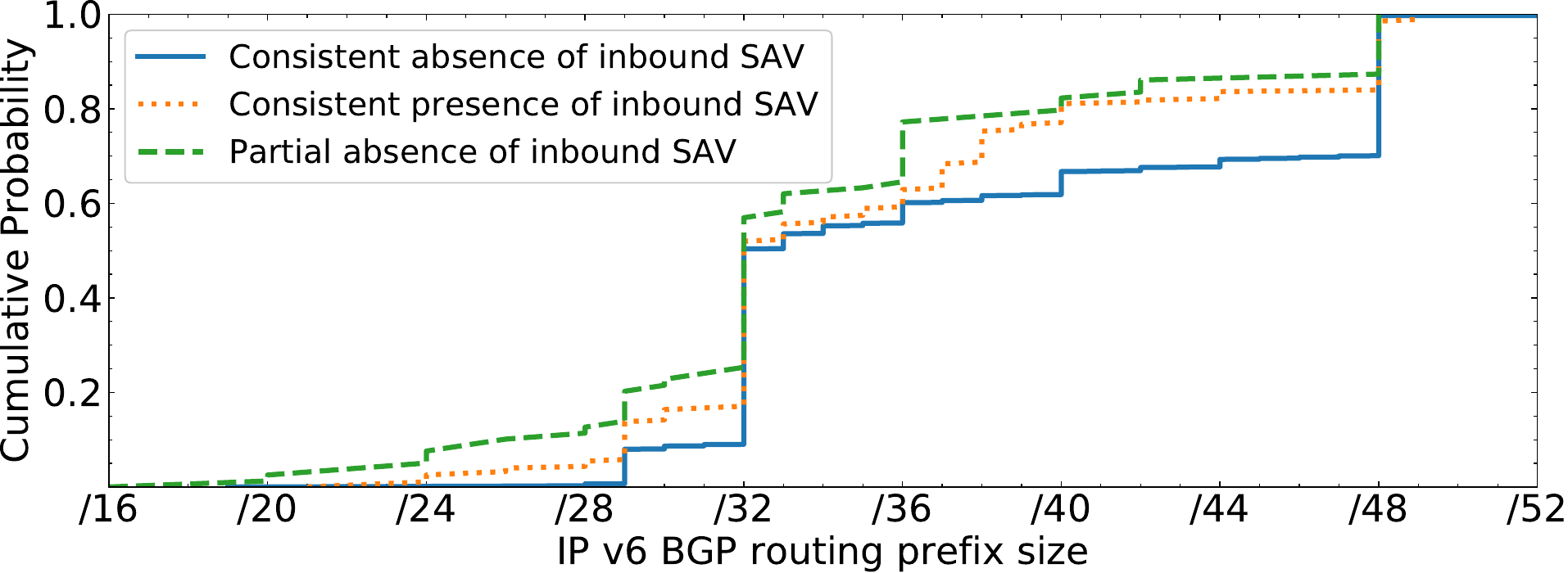}
  \caption{Sizes of the IPv6 longest matching prefixes from the BGP routing table. Prefixes with consistent absence of inbound SAV tend to be the smallest.}
  \label{cdf_prefix_sizes_v6}
\end{figure}

\section{Impact of Network Characteristics on SAV Policies\label{sec:network}}

Multiple factors may influence the decision of network operators to deploy (or not) inbound SAV in their networks. We explore some of potential reasons below.

\textit{1) Size of the address space:} Previous work assumed that the size plays an important role in SAV deployment for \textit{outbound} traffic: it is less likely that smaller organizations have resources and incentives to implement packet filtering in their networks \cite{loops,saving}. We need to consider this assumption with caution, as the two methods relied on scanning a sample of networks and the results are not necessarily representative for the general population. Yet, we hypothesize that operators with a larger address space are more likely to adhere to best current practices and promote routing security. For example, MANRS (Mutually Agreed Norms for Routing Security regulations)~\cite{manrs} members are strongly encouraged to implement SAV in edge routers. 

Figure~\ref{cdf_as_sizes} presents the cumulative distribution of IPv4 autonomous system sizes computed as the number of announced IPv4 addresses in the BGP routing table. It includes ASes of the three groups, as defined in Table~\ref{inconsistencies}: consistent absence, partial absence, and consistent presence of inbound SAV. The size distribution of consistently vulnerable ASes is driven by the small ones---as many as 75.3\% of these ASes contain 4,096 and fewer addresses. In contrast, partially vulnerable and non-vulnerable networks tend to be bigger---around half of them are /20 and bigger. We observe similar trends for the BGP prefix sizes (see Figures~\ref{cdf_prefix_sizes} and~\ref{cdf_prefix_sizes_v6}). Overall, our results show that non-compliant networks tend to be smaller than (partially) compliant. One possible explanation is that the operators of smaller networks have generally fewer resources and lower competence to deploy SAV in both directions.

\textit{2) AS stability:} It is challenging to implement ACL-based inbound SAV if BGP advertisements change frequently. We hypothesize that ASes with stable BGP advertisements are more likely to have SAV in place, as it would be easier to implement static ACLs. We define AS stability as the ratio of prefixes that remain the same compared to all announced prefixes over time. We analyze weekly BGP announcements \cite{routeviews} for the period between September 2019 and March 2020. We find that 87\% of ASes with consistent presence and 86\% of ASes with the consistent absence of inbound SAV advertise exactly the same prefixes. On the contrary, fewer ASes that partially deploy inbound SAV are stable (81\%). While there is a 5-6\% difference between ASes with consistent filtering and ASes with partial filtering, contrary to our hypothesis, AS stability does not seem to play a significant role in the decision of network operators to deploy SAV.

\textit{3) AS type:}  there are three types of autonomous systems with respect to the way they are interconnected: stub (connected to a single upstream provider), multihomed (connected to multiple upstream providers), and transit (interconnect other autonomous systems)~\cite{astypes}. The last two are commonly referred to as ``non-stub''. The problem with non-stub providers is that they might have customer ASes that do not announce all routes to them due to load balancing or fault tolerance~\cite{953233}. Similarly, strict filtering might not be feasible for non-stub ASes in the case of asymmetric routing, particularly for multi-homed networks~\cite{Baker:2004:IFM:RFC3704}.

We rely on the CAIDA AS relationship data~\cite{dimitropoulos2007relationships} to find the types of IPv4 autonomous systems that we have shown to be vulnerable or not. We find that 95\% and 90\% of ASes with consistent absence and presence of inbound SAV, respectively, are stub ASes. At the same time, fewer ASes with partial absence of inbound SAV (77\%) are stubs. As stubs connect to only one upstream provider, it is easier to implement SAV using, for example, static access control lists. 

\textit{4) Other:} We have approached the operators of 30 networks for which inbound SAV is partially deployed and asked their reasons for not deploying inbound SAV consistently. We got replies from three administrators. One administrator manages a /24 IPv4 network that is logically divided into two parts. Some IP addresses belong to virtual machines and their OpenStack configuration provides inbound and outbound SAV, while others are physical servers or Internet access subscribers that do not deploy inbound SAV due to complexity, time, and financial issues. Another network administrator confirmed being responsible only for a subset of the /24 IPv4 network, thus having no control over the other part.  Indeed, upstream providers may perform route aggregation of smaller customer networks, maintained by different organizations~\cite{loops} that possibly implement different anti-spoofing policies. Finally, one network operator reported that the whole /24 network had no inbound SAV in place, so we must have encountered packet losses.

\section{Outbound versus Inbound SAV Policies\label{sec:policies}}

Outbound SAV is the most effective way to prevent IP spoofing attacks at their origin. Although it protects the rest of the Internet, the deploying network does not directly benefit from it. On the contrary, inbound SAV protects the network itself from external attacks. We hypothesize that network operators are more incentivized to deploy inbound SAV than outbound. In this section, we analyze which type of SAV is deployed at the /24 IPv4 (/40 IPv6) network and autonomous system levels.

\subsection{Network Level} 

Our scans reveal \textit{inbound} filtering policies of 1.2M /24 IPv4 and 10.5K /40 IPv6 networks. To check whether these networks configure \textit{outbound} SAV, we apply the forwarder-based method proposed by Mauch~\cite{mauch} on our own data and refer to one external data source (the Spoofer project). We next look for the overlap between these data sources.

As explained in Section~\ref{sec:relatedSAV}, the Spoofer client works by sending spoofed and unspoofed packets using the installed client software. The outbound SAV results are anonymized per /24 IPv4 and /40 IPv6 address blocks and made publicly available. The Spoofer identifies four possible states: \textit{blocked} (only an unspoofed packet was received, the spoofed packet was blocked), \textit{rewritten} (the spoofed packet was received, but its source IP address was changed on the way), \textit{unknown} (neither packet was received), \textit{received} (the spoofed packet was received by the server). 

In March 2020, we collected and aggregated the latest Spoofer data for one month. We obtained the tests for 3,731 /24 IPv4 and 579 /40 IPv6 networks. We only keep vulnerable to spoofing (\textit{received}) and non-vulnerable to spoofing (\textit{blocked}) networks, leaving out the other two categories (\textit{rewritten} and \textit{unknown}) as non-conclusive. Note that these numbers are much smaller than 9,751 ASes tested by the Spoofer since 2015. For the comparison in this section, we choose only the tests conducted around the same period as our scans. The overlap between our inbound method and the Spoofer tests represents 473 /24 IPv4 and 17 /40 IPv6 networks. The minority of them have consistent filtering in both directions: 91 /24 IPv4 (3 /40 IPv6) networks have no filtering in both directions while 77 /24 IPv4 (2 /40 IPv6) networks implemented both inbound and outbound SAV. Interestingly, whenever filtering is deployed only in one direction, it is mostly outbound (59.4\% for IPv4 and 70.6\% for IPv6). 

We further compare two Spoofer datasets---outbound and inbound. The latter, as mentioned in Section~\ref{comparisonspoofer}, is not publicly available and was obtained by contacting CAIDA directly. We found that the Spoofer had results for both inbound and outbound tests for 298 /24 IPv4 and 273 /40 IPv6 networks. The majority of them were consistent in both directions (63\% and 65\%, respectively). The remaining networks (deploying SAV in only one direction) were mostly filtering outbound traffic (31\% for IPv4 and 32\% for IPv6).

The next outbound SAV dataset comes from the forwarder-based measurement technique. The overlap with the inbound SAV dataset is 16K IPv4 and 3 IPv6 networks. All the IPv6 networks had no SAV in both directions. For IPv4, 33.2\% of networks have no SAV in both directions, whereas most of the IPv4 networks without outbound SAV (66.8\%) deploy inbound SAV. The important limitation of the forwarder-based method consists of inability to identify the presence of outbound SAV. Therefore, we need to consider these results with caution because we cannot compare networks with deployed outbound SAV but without SAV for inbound traffic using the forwarder-based method and our scheme.

\subsection{Autonomous System Level} 

To evaluate the type of SAV (inbound or outbound) more deployed at the autonomous system level, we analyzed a subset of ASes, namely MANRS members~\cite{manrs}.

MANRS is one of the most well-known initiatives to improve the security and resilience of the Internet global routing system. It requires its members to adhere to a set of compulsory and recommended actions. ``Preventing traffic with spoofed source IP addresses'' falls into the recommended category, yet is highly encouraged. The compliance with these standards is observed by the MANRS Observatory~\cite{observatory}, a data collection system and an ``online tool to constantly monitor the state of Internet routing security.'' In particular, it tracks the number of networks ``preventing traffic with spoofed source IP addresses'' since 2019.

As the per-AS data is not available to non-members, we retrieved the list of participating autonomous systems (515 at the time of writing) to find the overlap between the Spoofer dataset, forwarder-based dataset, and our inbound SAV measurements. Recent work showed that MANRS members are not more likely to deploy SAV than the general population~\cite{spoofer_new}. We show that 81 MANRS ASes out of 515 are vulnerable to spoofing of outbound traffic (as shown by the Spoofer and the forwarder-based datasets), but as many as 311 ASes are at least partially vulnerable to spoofing of inbound traffic. 

Therefore, our results suggest that network operators are familiar with the concept of SAV but they tend to secure traffic only leaving their networks. This observation applies to small networks (/24 IPv4 and /40 IPv6) as well as to large autonomous systems.

\section{SAV Deployment for IPv4 and IPv6}\label{sec:deploymv4v6}

As IPv6 deployment is growing, it becomes an attractive attack target. Previous work showed that individual hosts as well as larger networks are generally more open over IPv6~\cite{back_door}. In this section, we analyze whether dual-stack autonomous systems and individual hosts are more protected from spoofing attacks over IPv4 than over IPv6.

\begin{table}[t]
    \caption{Fingerprinting dual-stack candidate pairs} 
    \label{fing} 
    \scriptsize
    \centering 
    \setlength{\tabcolsep}{6pt}
    \begin{tabular}{lccccc}
         \toprule
        \multirow{3}{5em}{\textbf{Protocol/ Application}} & 
        \multirow{3}{2em}{\textbf{Both closed}} & 
        \multirow{3}{2em}{\textbf{Only IPv4 open}} & 
        \multirow{3}{2em}{\textbf{Only IPv6 open}} & 
        \multirow{3}{2em}{\textbf{Both open}} & 
        \multirow{3}{5em}{\textbf{Same fingerprint}}  \\
         &&&&&\\ 
         &&&&&\\
         \midrule
         DNS (\texttt{\footnotesize{version.bind}}) & 16.7k & 13.1k & 1.7k & 50k & 37.3k (45.8\%)\\
         DNS (\texttt{\footnotesize{PTR}}) & 11.4k & 38.1k & 1.2k & 30.9k & 24k (29.4\%)\\
         NTP & 67.1k & 2k & 2.5k & 10k & 128 (0.2\%)\\
         HTTP & 27.4k & 16k & 3.3k & 35k & 34.2k (41.9\%)\\
         HTTPS & 29.1k & 16.8k & 675 & 35k & 22.5k (22.6\%)\\
         SSH & 33.8k  & 2k & 2.4k & 43.3k & 5.6k (6.9\%)\\
         SMTP & 47.6k & 10.1k & 653 & 23.2k & 23.1k (28.3\%)\\
         
         \midrule
         Total (unique) &&&&& 61.3k (75.2\%)\\
         \bottomrule
    \end{tabular}
\end{table}

\begin{table*}[t]
    \caption{Geolocation results} 
    \label{geolocation} 
    \scriptsize
    \centering 
    \begin{tabular}{llclclclclc}
         \toprule
        \multirow{4}{*}{\textbf{Rank}} & \multicolumn{4}{c}{\multirow{3}{*}{\textbf{Resolvers (\mypound)}}} & \multicolumn{4}{c}{\multirow{3}{*}{\textbf{Networks, vulnerable to spoofing of inbound traffic (\mypound)}}} & \multicolumn{2}{c}{\multirow{3}{15em}{\centering \textbf{Proportion of  networks, vulnerable to spoofing of inbound traffic (\%)}}} \\
        &&&&&&&&&&\\
        &&&&&&&&&&\\
        \cmidrule(lr){2-5}
        \cmidrule(lr){6-9}
        \cmidrule(lr){10-11}
        & {\textbf{Country}} & {\textbf{IPv4}} & {\textbf{Country}} & {\textbf{IPv6}}
        & {\textbf{Country}} & {\textbf{IPv4}} & {\textbf{Country}} & {\textbf{IPv6}}
        & {\textbf{Country}} & {\textbf{IPv4}} \\
         \midrule
         1 & China & 1,970,410 & USA & 22,992 & China & 260,047 & USA & 1,319 & Kosovo & 63.6 \\
         2 & Brazil & 667,036 & Germany & 13,373 & USA & 162,259 & Brazil & 930 & Comoros & 52.6 \\
         3 & USA & 661,943 & Netherlands & 11,514 & Russia & 54,451 & Germany & 680 & Western Sahara & 50.0 \\
         4 & Iran & 404,134 & Belarus & 7,455 & Italy & 32,026 & Netherlands & 336 & Armenia & 49.5 \\
         5 & India & 348,491 & Russia & 6,410 & Brazil & 28,836 & United Kingdom & 309 & Maldives & 39.7 \\
         6 & Algeria & 249,931 & China & 5,840 & Japan & 27,890 & China & 304 & Moldova & 38.2 \\
         7 & Russia & 224,985 & United Kingdom & 5,151 & India & 27,426 & Russia & 289 & Niue & 37.5 \\
         8 & Indonesia & 222,602 & Spain & 3,996 & Mexico & 23,288 & Czech Republic & 254 & Palestine & 36.3 \\
         9 & Italy & 105,476 & Czech Republic & 3,357 & United Kingdom & 16,976 & France & 223 & Afganistan & 36.2 \\
         10 & Argentina & 104,850 & France & 2,837 & Indonesia & 16,798 & Japan & 183 & Bulgaria & 36.0 \\
         \bottomrule
    \end{tabular}
\end{table*}

\subsection{Individual Host Level}

We requested all the IPv4 and IPv6 non-forwarders to resolve domain names that require changing the IP protocol version, e.g., from IPv4 to IPv6 and from IPv6 to IPv4. Out of 2.6M IPv4 (36K IPv6) resolvers, 2.7\% and 28.5\% had IPv6 and IPv4 connectivity, respectively. It is not surprising that IPv6 resolvers are much more accessible over IPv4 than the other way round. As the IPv6 adoption is far from universal~\cite{CzyzAZIOB14,NikkhahG16,LivadariuED17}, it is crucial for IPv6 resolvers to be reachable over IPv4.

We collected 82K candidate address pairs in total, most of them (72K) during the IPv4 scan. Indeed, DNS resolvers are known to have complex relationships and a single address may appear in multiple address pairs~\cite{nameserver}. However, for our analysis, we consider each address pair separately. We collect fingerprinting information for each address in the pair as described in Section~\ref{sec:fingerprint}. The great majority of the candidate pairs (98.1\%) had at least one fingerprinting port open for both IPv4 and IPv6, mostly DNS and SSH. Table~\ref{fing} presents the detailed results. These two ports are necessary for DNS resolvers to be accessible and to function properly. While the NTP port is also relatively open, in most cases, we could merely extract the timestamp. Only 128 server pairs returned the same software and operating system versions. Overall, 75.2\% of address pairs suitable for fingerprinting had matching signatures on at least one protocol/application. Two of the three network operators that responded to our survey operate dual-stack resolvers and they confirmed the correctness of our mappings. In particular, 6 pairs had identical \texttt{PTR} records and 7 pairs had identical \texttt{\small{version.bind}} records. The remaining pairs had either no record at all or only records for one address in the pair. Thus, we did not consider those pairs for classification.

From 61K seemingly dual-stack pairs, 43K responded to our spoofed and non-spoofed queries, revealing the absence or the presence of SAV for IPv4 and IPv6. Most of them (99.2\%) have consistent filtering policies. Out of the remaining 324 hosts, 195 (60.2\%) are vulnerable to spoofing of inbound traffic only over IPv6. Thus, at the individual host level, SAV tends to be consistently deployed for IPv4 and IPv6.

\subsection{Autonomous System Level}

Whenever a certain security policy exists for an individual dual-stack host, it is likely to hold for the whole autonomous system~\cite{back_door}. Consequently, we generally expect to have similar security practices for IPv4 and IPv6, because the great majority of networks have consistent policies for IPv4 and IPv6 at the host level. As of March 2020, there are 66,978 IPv4 and 18,710 IPv6 ASNs present in BGP routing tables. 18,016 of them advertised both IPv4 and IPv6 prefixes.

For this analysis, we choose vulnerable and non-vulnerable to inbound spoofing ASes and keep those having results for both IPv4 and IPv6. The resulting set includes 2,873 ASes. The great majority of them (94.2\%) have consistent filtering policies for IPv4 and IPv6---2,650  are vulnerable and 55 are non-vulnerable to inbound spoofing. As for the remaining 168 ASes, 19 (11.3\%) are only vulnerable for IPv4 and 149 (88.7\%) are only vulnerable for IPv6. Thus, we conclude that at the AS level, SAV for inbound traffic is generally deployed consistently for IPv4 and IPv6.

\section{Geographic Distribution\label{sec:geo}}

Identifying countries that do not comply with the SAV standard is the first step in mitigating the issue by, for example, contacting local CSIRTs. We use the MaxMind database \footnote{\url{https://dev.maxmind.com/
geoip/geoip2/geolite2/}}
to map every IP address encoded in the domain name (the original destination of the query) to its country. Table~\ref{geolocation} summarizes the results.

Overall, 232 countries and territories contain networks that do not deploy inbound SAV neither for IPv4, IPv6, nor for both. When counting DNS resolvers that responded to spoofed requests, most of them originate from China for IPv4 and from the USA for IPv6. As explained in Section~\ref{sec:res_v6}, the coverage of the IPv6 scan is smaller than that of IPv4, which is why we have found much fewer DNS resolvers. The top 10 ranking differs greatly for IPv4 and IPv6, as only three countries are present for both. We map individual resolvers to the corresponding /24 IPv4 and /40 IPv6 networks and aggregate them per country. Multiple resolvers are distributed in fewer networks, resulting in the updated top 10 ranking, different from the one for the resolver count.

\begin{figure}[t!]
\centering
\includegraphics[width=\linewidth]{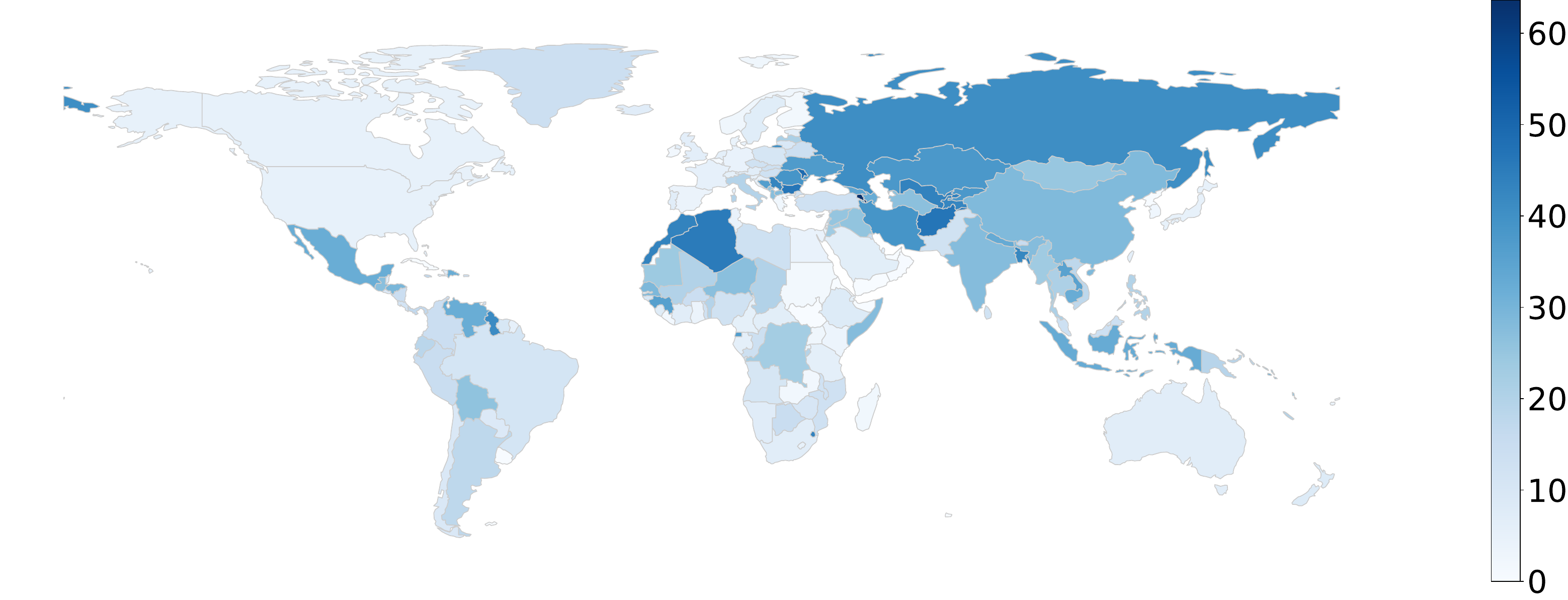}
\caption{Fraction of vulnerable to spoofing of inbound traffic vs. all /24 IPv4 networks per country (in \%)}
\label{spoofers_map}
\end{figure}

Such absolute numbers are still not representative as countries with a large Internet infrastructure may have many DNS resolvers and therefore reveal many vulnerable to spoofing of inbound traffic networks that represent a small proportion of the whole. For this reason, we compute the fraction of vulnerable to spoofing of inbound traffic vs. all /24 IPv4 networks per country. To determine the number of all the /24 networks per country, we map all the individual IPv4 addresses from the BGP routing table to their location, then to the /24 block, and keep the country/territory to which most addresses of a given network belong. Figure~\ref{spoofers_map} presents the resulting world map. We can see in Table \ref{geolocation} that the top 10 ranking has changed once again. Small countries such as Western Sahara and Niue that have two and eight identified resolvers each suffer from a high proportion of vulnerable to spoofing of inbound traffic networks. One of the two /24 networks of Western Sahara allows spoofing of inbound traffic. On the other hand, Bulgaria is a country with a large Internet infrastructure (16,439 /24 networks in total) and with a large percentage of vulnerable to spoofing of inbound traffic networks.

\section{Discussion\label{discussion}}

\subsection{Limitations\label{limitations}}

Our approach has some limitations that may impact the accuracy of the results. We rely on the main assumption---the presence of a DNS resolver (open or closed) in a tested network. If it is not present, we cannot conclude on the filtering policies. Closed DNS resolvers only reveal the absence of inbound SAV, at least for some part of networks they belong to. Open resolvers, on the contrary, reveal both the absence and the presence of inbound SAV (assuming that transit networks do not deploy SAV).

Transit filtering may drop our scan packets before they even reach the target networks. In some cases, we would not detect vulnerable networks (if only closed resolvers are present), in other cases, we would incorrectly classify vulnerable networks as non-vulnerable (if open resolvers are present). However, if our spoofed probes do arrive in the target network, we can detect the absence of inbound SAV. In this sense, our results indicate the lower bound of the problem---in an ideal measurement setup without SAV in transit networks, we could detect a larger number of networks vulnerable to spoofing of inbound traffic.

Some other reasons, such as packet losses or temporary network failures, may also explain the absence of data for certain IP addresses.

\subsection{Ethical Considerations\label{ethics}}

To make sure that our study follows the ethical rules of network scanning, yet providing complete results, we have adopted the recommended best practices~\cite{eth,zmap}. For the IPv4 scan, we aggregate the BGP routing table to eliminate overlapping prefixes. In this way, we send no more than two DNS \texttt{A} request packets (spoofed and non-spoofed) to every tested host. Due to packet losses, we potentially miss some results, but we accept this limitation not to disrupt the normal operation of tested networks. In addition, we randomize our input list for the scanner so that we do not send consecutive requests to the same network (apart from two consecutive spoofed and non-spoofed packets). We spread our scanning activities over 15 days due to limited resources on the scanning machine (8 vCPUs and 3GB of RAM).

We have set up a website for this project on \texttt{\small{\url{https://closedresolver.korlabs.io}}} and provided all the queried domains and the fingerprinting server with a description of our project as well as the contact information if someone wants to exclude her networks from testing. We have received 9 requests from operators of, among others, /8, /9, and /10 IPv4 networks, who noticed our DNS requests. In total, we excluded 29M IPv4 addresses from futures scans as well as two IPv6 prefixes (/128 and /48). We also exclude these addresses from our analysis. We do not publicly reveal the SAV policies of individual networks and AS operators. Yet, website visitors can see the results for the network they connect from. 

\section{Conclusions}\label{sec:conclusion}

In this paper, we have presented a novel method to infer the deployment of inbound SAV for the IPv4 and IPv6 address spaces. We have measured the filtering policies of 52\% of routable IPv4 ASes 
(26\% for IPv6) and 28\% of all the IPv4 BGP prefixes (almost 9\% for IPv6). We show that most 
of the networks for which we obtained measurements are consistently or partially vulnerable to spoofing of inbound~traffic.

Reflection DDoS attacks have extensively used open DNS resolvers in recent years. We have found 5.3M IPv4 and 16K IPv6 open resolvers. New ways to misuse open resolvers constantly emerge. For example, NXNSAttack can exploit open recursive resolvers to reach an amplification factor of up to 1,620. Even worse, spoofing of inbound traffic combined with the NXNSAttack results in additional 2.5M closed resolvers for IPv4 (100K for IPv6) that might be either vulnerable themselves or possibly misused against other victims.

Open resolvers when they do not resolve spoofed queries identify the presence of inbound SAV at the edge of the tested network or filtering in transit. We found that while many providers deploy consistent filtering policies network-wide, there are cases when a single network is only partially protected from spoofing of inbound traffic. The results indicate that different network characteristics are factors that prevent operators from correctly configuring packet filtering. Overall, the proportion of non-vulnerable networks is much lower compared to networks with the consistent or partial absence of inbound SAV.

We have identified and fingerprinted dual-stack DNS resolvers and shown that at the individual host level, inbound filtering is generally deployed consistently for IPv4 and IPv6. In the remaining few cases, the IPv4 part is more secure than IPv6. This observation also holds for dual-stack ASes. 

We have gathered different datasets to analyze whether outbound filtering is less deployed than inbound. Outbound SAV faces the problem of misaligned economic incentives---it protects other networks but not the one deploying it. Interestingly, SAV for outbound traffic turned out to be more deployed than inbound at the AS level among network operators committed to the MANRS initiative. The absence of outbound packet filtering gained widespread attention since it enables DDoS attacks. Under these circumstances, inbound SAV remains neglected (or overlooked) by network operators.

Vulnerability to spoofing of inbound traffic is not limited to any geographic territory and is spread worldwide. To draw attention to the problem of spoofing of inbound traffic, we launched the Closed Resolver Project at \texttt{\small{\url{https://closedresolver.korlabs.io}}}.
Anyone can visit the project website and check whether his/her network is vulnerable to spoofing of inbound traffic and how many closed resolvers we found inside. The long-term objective is to 
run notification campaigns for network operators and provide them 
with an accessible platform to investigate results for their networks. The service is particularly useful for operators planning to become MANRS participants since MANRS strongly recommends deploying SAV. We expect these efforts will result in better packet filtering on the Internet.

\section*{Acknowledgements}
The authors would like to thank  the reviewers and the editor for their valuable 
feedback. This work was partially supported by RIPE NCC, Carnot LSI, Grenoble Alpes Cybersecurity Institute (under the contract ANR-15-IDEX-02), PERSYVAL-Lab project (under the contract ANR-11-LABX-0025-01), and DiNS project (under the contract ANR-19-CE25-0009-01).



\bibliographystyle{IEEEtran}
\bibliography{references}

\begin{IEEEbiography}[{\includegraphics[width=1in,height=1.25in,clip,keepaspectratio]{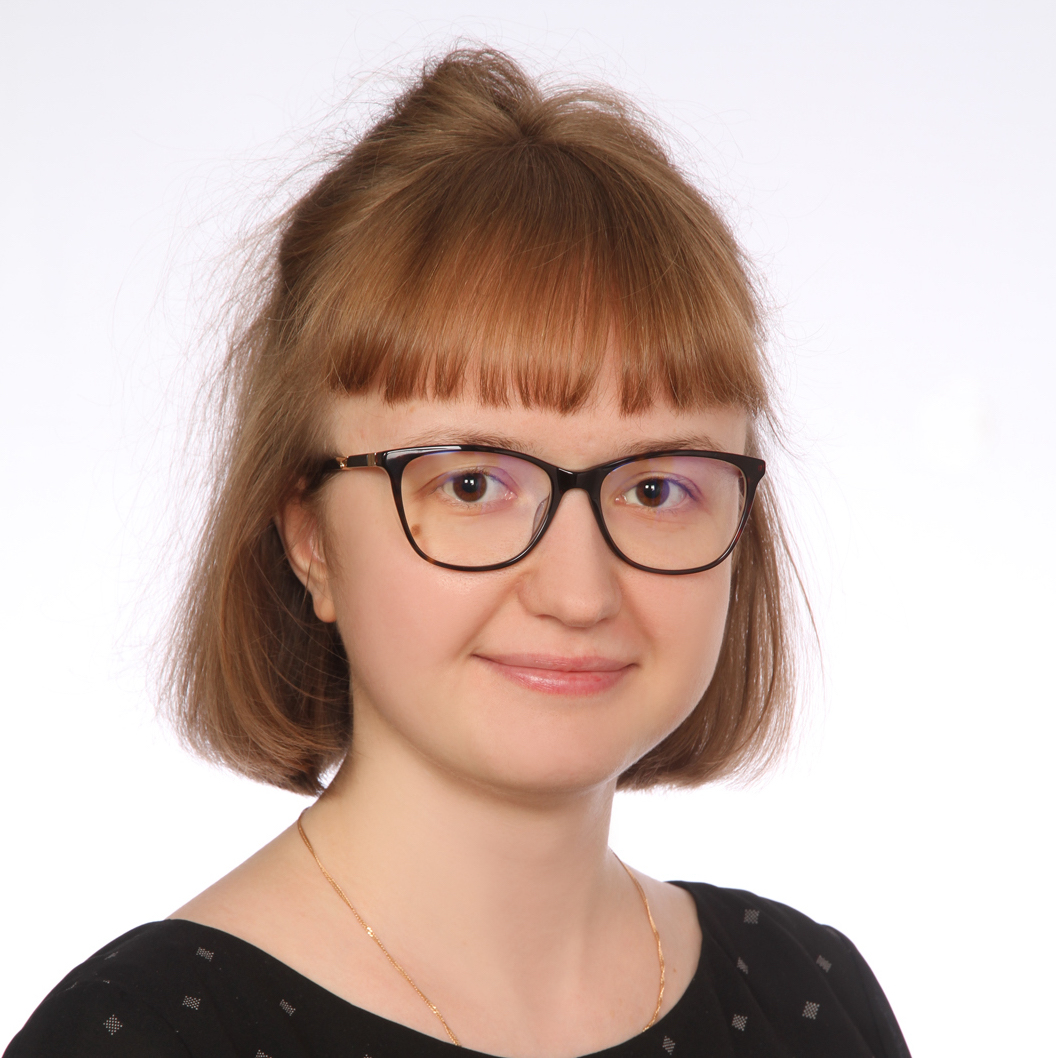}}]{Yevheniya Nosyk} 
is a Ph.D. student at Université Grenoble Alpes, France. She received her Bachelor's degree in Information Technology from the South-Eastern Finland University of Applied Sciences (2018) and a Master's degree in Computer Science from the Université Grenoble Alpes (2021). She is currently doing research in the area of network security and DNS from a large-scale Internet measurements point of view.
\end{IEEEbiography}

\begin{IEEEbiography}[{\includegraphics[width=1in,height=1.25in,clip,keepaspectratio]{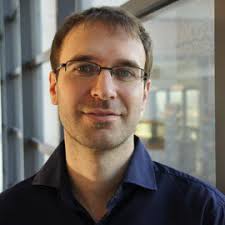}}]{Maciej Korczy\'nski} 
is an Associate Professor at Grenoble Institute  of  Technology, France. He received the HDR (2021) and Ph.D. (2012) degrees in computer science from the Université Grenoble Alpes. He was a post-doctoral researcher at the Rutgers University, USA (2013-2014) and Delft University of Technology, the Netherlands (2014-2017). His main research interests include Internet-wide passive and active  measurements for cybersecurity, domain name abuse, incident data analysis, vulnerability notifications, economics of cybersecurity, and security of Internet protocols, with a~focus~on~DNS.
\end{IEEEbiography}

\begin{IEEEbiography}[{\includegraphics[width=1in,height=1.25in,clip,keepaspectratio]{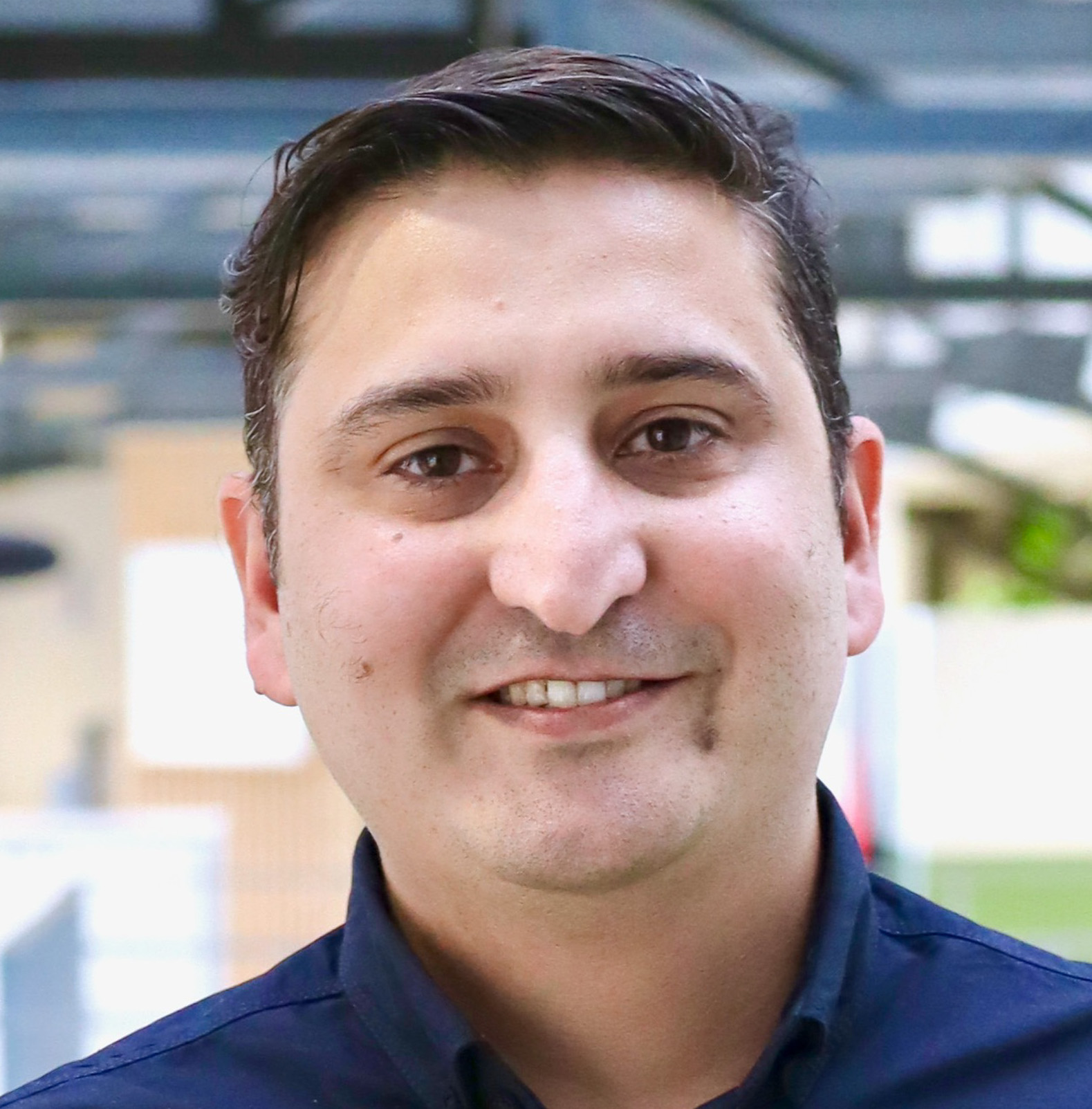}}]{Qasim Lone} 
is a senior research engineer in the R\&D department at RIPE NCC. He received his Ph.D. from the Delft University of Technology, The Netherlands. Previously, he has also worked as a visiting scientist at SLAC National Accelerator Laboratory (SLAC), Stanford University, USA. His research interests include internet measurements, Internet outages, data analysis, and cybersecurity. 
\end{IEEEbiography}

\begin{IEEEbiography}[{\includegraphics[width=1in,height=1.25in,clip,keepaspectratio]{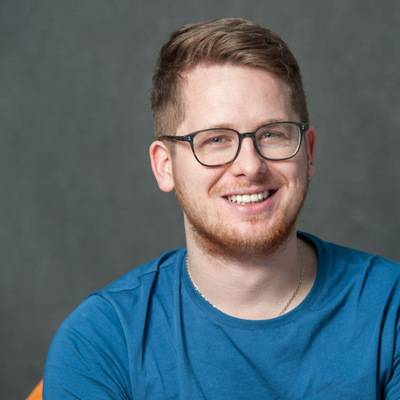}}]{Marcin Skwarek} 
received both his Master's and Engineer's telecommunication degrees from Warsaw University of Technology. He works as Senior R\&D Software Engineer at Exatel where he develops highly efficient, secure, and reliable telecommunication software. He occasionally does research related to internet-wide measurements, software deployments and misconfigurations, network traffic approximation, NoSQL databases, and anything connected to DNS protocol. He is committed to providing suitable solutions to challenging problems in professional and academic areas. Personally a big fan of free software especially GNU Emacs and Linux.
\end{IEEEbiography}

\begin{IEEEbiography}[{\includegraphics[width=1in,height=1.25in,clip,keepaspectratio]{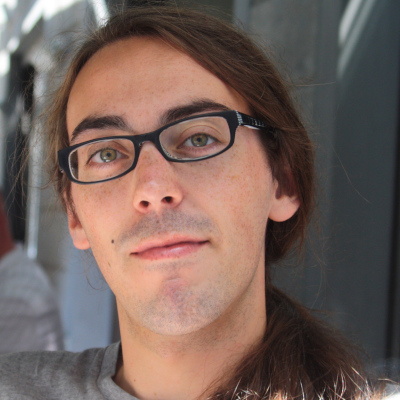}}]{Baptiste Jonglez} 
is a Research Engineer at Inria, the French national institute for research in digital science and technology.  He is currently working in the STACK research group, where he is the architect of the team's software contributions in Edge Computing.  He received his Ph.D. in Computer Science from Université Grenoble Alpes, France, in 2020. His research interests encompass networks and systems in the context of the Internet. He has a specific focus on experimental work and open large-scale research platforms and more generally strives to make sure that theory and practice can meet to solve the challenges arising from modern network and system infrastructures.
\end{IEEEbiography}

\begin{IEEEbiography}[{\includegraphics[width=1in,height=1.25in,clip,keepaspectratio]{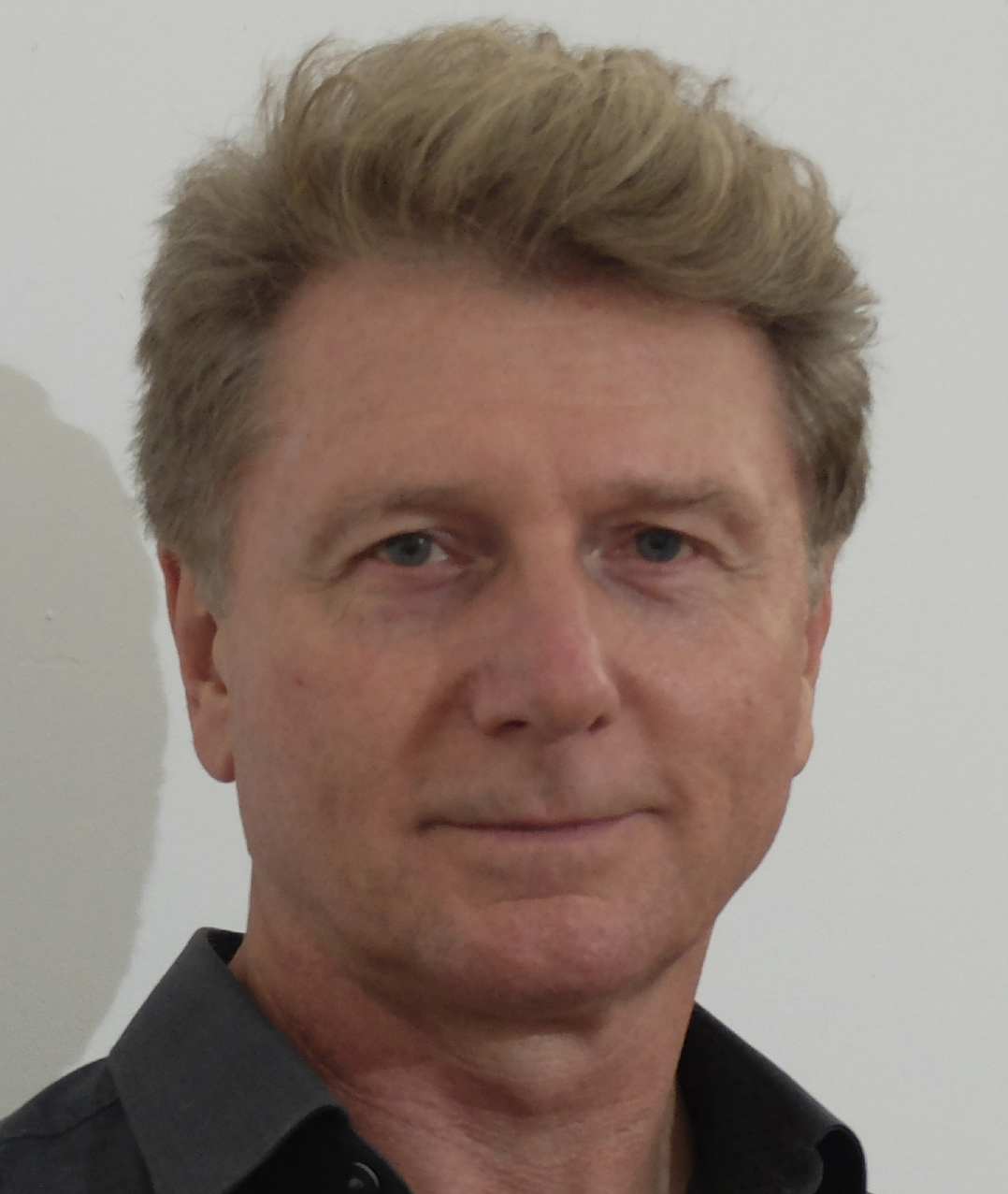}}]{Andrzej Duda}
is a Full Professor at Grenoble Institute of Technology. He received his Ph.D. from the Université de Paris-Sud and his Habilitation diploma from  Grenoble University. Previously, he was an Assistant Professor at the Université de Paris-Sud, a Chargé de Recherche at CNRS, and a Visiting Scientist at the MIT Laboratory for Computer Science. In 2002-2003, he was an Invited Professor at EPFL (Swiss Federal Institute of Technology in Lausanne). He published over 180 papers in the areas of performance evaluation, distributed systems, multimedia, and networks. 
\end{IEEEbiography}

\end{document}